\documentclass[lettersize,journal]{IEEEtran}
\usepackage{amsmath,amsfonts}
\usepackage{algorithmic}
\usepackage{algorithm}
\usepackage{array}
\usepackage[caption=false,font=normalsize,labelfont=sf,textfont=sf]{subfig}
\usepackage{textcomp}
\usepackage{stfloats}
\usepackage{url}
\usepackage{verbatim}
\usepackage{graphicx}
\usepackage{cite}
\usepackage{multirow}
\usepackage[table]{xcolor}

\begin{document}

\title{Accelerating Network-Agent Dispersion:\\
Territorial Behavior and Directionally Biased Lazy Random Walks}

\author{Li Zeng, Shenzhen Polytechnic University, Guangdong, China\\Steve Alpern, Warwick Business School, University of Warwick, Coventry, UK}



\maketitle

\begin{abstract}
Territorial behavior can greatly accelerate decentralized agent dispersion on networks. This paper studies a network-agent dispersion problem in which $m$ autonomous agents move in discrete time on a connected graph and seek a configuration in which no two agents occupy the same node. We focus on the dispersion case $m=n$, where successful configurations contain exactly one agent per node. In the baseline model, each agent follows a lazy random walk with a common laziness parameter $p$. This process defines a finite absorbing Markov chain, and the expected absorption time is used to measure dispersion efficiency. We introduce two local behavioral extensions: territorial behavior, in which an agent that is alone at a node claims that node and repels later arrivals, and directional bias, in which agents share a preferred direction of movement on paths and cycles. Exact calculations on three-agent path and cycle networks and Monte Carlo simulations on larger instances show that territorial behavior substantially reduces expected dispersion time, with larger relative reductions as network size increases. Directional bias alone has limited effect in most small-network cases, but when combined with territorial behavior it can produce large additional speedups. In particular, the simulations show reductions of $99.22\%$ on $L_{100}$ and $97.48\%$ on $C_{100}$ when all agents start from one node. These results show how simple local movement rules can strongly affect global dispersion time in decentralized networked multi-agent systems.
\end{abstract}


\section{Introduction}
\IEEEPARstart{D}{ecentralized} mobile agents often need to spread across a network without being assigned final positions by a central controller. Robotic teams, unmanned aerial vehicle swarms, sensor systems, and emergency-response units may need to avoid crowding, reduce interference, or provide spatial coverage while relying only on simple local movement rules. This paper studies how such rules affect the time required for agents to reach a dispersed configuration, and shows that a simple territorial rule can greatly accelerate this process.

We consider a network-agent dispersion problem in which agents move in discrete time on a connected graph. The main case studied in this paper is the dispersion setting $m=n$ and $D=1$, where the objective is to reach a configuration in which every node is occupied by exactly one agent. The baseline process is a lazy random walk with a common laziness parameter $p$. This gives a finite absorbing Markov chain, where the absorbing states are the dispersed configurations and the expected absorption time measures dispersion efficiency.

The main finding is that territorial behavior is the strongest acceleration mechanism. Under territorial behavior, an agent that is alone at a node claims that node and forces later arrivals to move away. This rule stabilizes partial progress toward dispersion because claimed nodes are no longer repeatedly lost to crowding. Directional bias alone is usually much less effective, but it becomes useful when combined with territorial behavior. Once ownership creates a stable occupied region, directional bias can guide remaining non-owner agents toward unclaimed nodes. In the simulations, this combined mechanism reduces expected dispersion time by $99.22\%$ on $L_{100}$ and by $97.48\%$ on $C_{100}$ when all agents initially start from one node.

A simple motivating example is a road network after a power failure, when traffic lights are no longer operating. Traffic officers may need to spread across intersections, but each officer may know only the local road layout and may not know where the others are. A global signal can indicate when full coverage has been achieved, without assigning each officer a destination. Similar issues arise in robotic and unmanned aerial vehicle deployment, where agents may need to separate across a network while using limited local information.

The work is related to the broader literature on networked multi-agent systems and collective behavior. Consensus, cooperation, and nearest-neighbor coordination provide central models for distributed control and robotic networks~\cite{OlfatiSaberFaxMurray2007,JadbabaieLinMorse2003,BulloCortesMartinez2009}. Swarm and self-driven particle models also show how simple individual rules can generate collective spatial organization~\cite{VicsekCzirokBenJacobCohenShochet1995,BonabeauDorigoTheraulaz1999}. The objective considered here is different from consensus or formation control: agents do not seek agreement or a prescribed formation, but instead must separate across the nodes of a graph through stochastic local movement.

The closest line of work concerns dispersion, social distancing, and related games on networks. Alpern and Reyniers introduced spatial dispersion as a dynamic coordination problem for non-communicating agents~\cite{AlpernReyniers2002}. Dispersion games and decentralized learning models were later studied from learning and economic perspectives~\cite{GrenagerPowersShoham2002,BlumeFranco2007}. Alpern and Zeng formulated social distancing, gathering, and search games for mobile agents on simple networks~\cite{AlpernZeng2022}. A preliminary version of the three-agent dispersion setting considered here was presented in~\cite{AlpernZeng2023}. Social-distance games have also been analyzed in terms of non-cooperative behavior and Pareto efficiency~\cite{BalliuFlamminiMelideoOlivetti2019,BalliuFlamminiMelideoOlivetti2022}. Compared with these game-theoretic or equilibrium-based formulations, the present paper emphasizes the expected time for a stochastic movement process to reach a dispersed absorbing state.

The model also connects to the broader literature on stochastic movement, diffusion, and spreading processes on networks. Random walks and diffusion provide standard tools for describing movement and propagation on graphs~\cite{MasudaPorterLambiotte2017,RiascosMateos2021}. Random-walk-based methods have also been used in network embedding and related network analysis tasks~\cite{LinSussmanIshwar2023}. Related network-science studies have examined spreading dynamics in epidemic, temporal, and multiplex network settings~\cite{JainBhatnagarPrasadKaur2023,FrieswijkZinoCao2023,ZhaoLiWangPeng2023}. Although the objective here is dispersion rather than contagion or feature diffusion, these works illustrate how local transition rules can shape network-level outcomes.

Territorial behavior provides the first behavioral modification of the baseline random walk. The idea is motivated by ecological models of habitat use, resource defense, and territory formation. Territorial behavior has long been used to explain spatial distributions in animal populations~\cite{FretwellLucas1969}, and later models have studied how local interactions can create persistent territorial structures~\cite{GiuggioliPottsHarris2011,WhiteVandeWoudeCraft2020}. Models of animal aggregation and dispersal also show how local behavioral rules can influence spatial separation~\cite{BroomErovenkoRowellRychtar2020}. In the present paper, territoriality is deliberately simplified into a local exclusion mechanism: a lone agent claims its node, remains there, and prevents later arrivals from staying.

Directional bias provides the second behavioral modification. In an unbiased lazy random walk, a moving agent chooses uniformly among neighboring nodes. Under directional bias, agents instead share a preferred direction of movement. Biased random walks have been used to model movement and search processes in structured environments~\cite{BonaventuraNicosiaLatora2014}. In this paper, directional bias is analyzed on paths and cycles, where the direction corresponds to left/right or clockwise/counterclockwise movement. More generally, such a rule would require additional graph structure, such as an orientation, a spatial embedding, or a distinguished target node. Thus, directional bias is treated as a structured extension of the lazy random walk rather than as a universal rule for arbitrary graphs.

Search and patrolling games provide another related direction, especially when agents or external authorities must locate, monitor, or redirect mobile entities on networks~\cite{AlpernMortonPapadaki2011,DuvocelleFleschStaudiglVermeulen2022,BasilicoDeNittisGatti2017}. These problems differ from the cooperative dispersion model studied here, but they suggest natural extensions involving external enforcement, strategic behavior, or moving targets.

The contributions of this paper are as follows:
\begin{itemize}
\item We formulate network-agent dispersion under lazy random walks as a finite absorbing Markov chain and use expected absorption time as the measure of dispersion efficiency.
\item We introduce territorial behavior and directionally biased lazy random walks as two local behavioral extensions of the baseline process.
\item We derive exact expected dispersion times for three-agent path and cycle networks and compare optimal movement parameters under different behavioral assumptions.
\item We use Monte Carlo simulations to show that territorial behavior substantially reduces expected dispersion time on larger paths and cycles, with relative improvements increasing as the network size grows.
\item We show that directional bias alone has limited effect in most small-network cases, but can produce large additional acceleration when combined with territorial behavior, especially when ownership creates a meaningful direction of progress.
\end{itemize}

The remainder of the paper is organized as follows. Section~II introduces the dispersion model, including lazy random walks, territorial behavior, directional bias, and the absorbing Markov chain formulation. Section~III presents exact calculations for three-agent dispersion on small path and cycle networks. Section~IV reports Monte Carlo simulation results for larger networks and compares the effects of the different movement rules. Section~V concludes and discusses possible extensions.

\section{Model and Notation}
\label{sec:model}

We consider a connected graph $Q=(V,E)$ with $n$ nodes and $m$ mobile agents, where $1\leq m\leq n$. All edges have unit length, and distances are measured by graph distance, that is, the number of edges in a shortest path between two nodes. Time is discrete, and at each time step every agent occupies one node of $Q$.

\subsection{States and dispersion objective}

In the non-territorial model, a state records the number of agents at each node. We write a state as
\[
s=[j_1,j_2,\dots,j_n],
\]
where $j_i$ is the number of agents at node $i$ and
\[
\sum_{i=1}^{n}j_i=m.
\]
The value $j_i$ is called the population of node $i$.

For a state $s$, let $d(s)$ be the minimum graph distance between any two distinct agents, where two agents occupying the same node have distance zero. For a fixed parameter $D\geq 1$, a state is called $D$-separated if $d(s)\geq D$.

The general $D$-separation problem on $(m,D,Q)$ is to understand how the agents move so that the process reaches a $D$-separated state and to quantify the expected time required.

This paper focuses on the dispersion case $D=1,\ m=n$.
When $D=1$, the condition $d(s)\geq 1$ means that no two agents occupy the same node. Since the number of agents equals the number of nodes, every dispersed state contains exactly one agent at each node:
\[
j_i=1,\qquad i=1,\dots,n.
\]
This is the spatial dispersion setting related to the model of Alpern and Reyniers~\cite{AlpernReyniers2002} and to later work on social distancing, gathering, and search games on simple networks~\cite{AlpernZeng2022}.

More generally, not every triple $(m,D,Q)$ admits a $D$-separated state. Let $\alpha_D(Q)$ denote the maximum number of nodes in $Q$ whose pairwise graph distances are all at least $D$. Under this convention,
\[
\alpha_1(Q)=|V|,
\]
while
$\alpha_2(Q)$ is the usual independence number of $Q$. A $D$-separated state exists only if
$m\leq \alpha_D(Q)$. 

In the present paper, because we focus on $D=1$ and $m=n=|V|$, this feasibility condition is automatically satisfied.

\subsection{Lazy random walks}

Random walks are widely used to model stochastic movement, exploration, and diffusion on networks~\cite{MasudaPorterLambiotte2017,RiascosMateos2021}. Random-walk-based methods also appear in network embedding and related network analysis tasks~\cite{LinSussmanIshwar2023}. The baseline motion rule in this paper is a lazy random walk with a common laziness parameter $p\in[0,1]$. Let
\[
q=1-p.
\]
When an agent is at a node $i$ with degree $\delta(i)$, it remains at $i$ with probability $p$ and moves to a neighbor of $i$ with total probability $q$. Conditional on moving, each neighbor is selected with probability
\[
\frac{q}{\delta(i)}.
\]
The parameter $p$ is called the laziness parameter, while $q$ is the movement probability. If $p=0$, the motion reduces to the usual simple random walk. If $Q$ is regular with degree $\delta$, then the choice
\[
p=\frac{1}{\delta+1}
\]
corresponds to adding a self-loop at each node and choosing uniformly among the node and its neighbors. This is often called a loop random walk.

In the non-territorial baseline model, agents move independently and do not use information about the positions of other agents. Therefore, under the baseline rule, the population vector evolves as a finite Markov chain.

\subsection{Directionally biased lazy random walks}

We also consider a directionally biased lazy random walk. Biased random walks extend the standard random-walk framework by allowing transition probabilities to depend on local preferences or graph structure~\cite{BonaventuraNicosiaLatora2014}. In this paper, directional bias is analyzed on paths and cycles, where a direction is naturally represented by left/right or clockwise/counterclockwise movement.

The directionally biased lazy random walk is described by parameters $p$, $r$, and $z$, where
\[
p+r+z=1,\qquad p,r,z\geq 0.
\]
At an interior node of a path, or at any node of a cycle, an agent remains at its current node with probability $p$, moves in the preferred direction with probability $r$, and moves in the opposite direction with probability $z$. When $r=z$, the rule reduces to the unbiased lazy random walk on paths and cycles. When $r\neq z$, agents have a shared directional preference.

At an endpoint of a path, there is only one neighboring node. In that case, an agent remains at the endpoint with probability $p$ and moves to the unique neighboring node with probability $1-p$. Thus, directional bias affects movement at interior nodes of paths and at all nodes of cycles, while endpoint movement is determined only by the laziness parameter.

\subsection{Territorial behavior}

Territorial behavior introduces a simple local exclusion mechanism. Nodes may be unowned or owned, and each node can be owned by at most one agent. Once an agent owns a node, it remains there and no longer follows the ordinary random walk. Other agents may arrive at an owned node, but they are not allowed to remain there permanently.

Because territorial behavior involves ownership, the population vector alone is not sufficient to describe the state of the process. For the territorial model, we use an augmented state
\[
s=([j_1,\dots,j_n],[o_1,\dots,o_n]),
\]
where $j_i$ is the number of agents at node $i$ and
\[
o_i=
\begin{cases}
1, & \text{if node } i \text{ is owned},\\
0, & \text{if node } i \text{ is unowned}.
\end{cases}
\]
Admissible territorial states satisfy
\[
j_i\geq o_i,\qquad i=1,\dots,n.
\]
If $o_i=1$, then one of the agents at node $i$ is the owner and remains fixed at that node. The number of non-owner agents at node $i$ is therefore $j_i-o_i$.

The territorial dynamics follow the timing convention below. At each time step, all owners remain fixed, and all non-owner agents move simultaneously. Non-owner agents that are not on owned nodes follow either the lazy random walk or the directionally biased lazy random walk, depending on the model being considered. If a non-owner agent is located on an owned node at the beginning of a movement step, it is forced to leave that node during that step. After all non-owner agents have moved, ownership is updated: any unowned node containing exactly one agent becomes owned by that agent.

Thus, the territorial rule can be summarized as follows:
\begin{itemize}
\item If an unowned node contains exactly one agent after movement, that agent claims the node.
\item An owner remains at its owned node with probability one.
\item If one or more non-owner agents occupy an owned node, each such non-owner agent is forced to leave at the next movement step.
\item Non-owner agents that are not forced to leave follow the applicable lazy or directionally biased movement rule.
\end{itemize}

When a non-owner agent is forced to leave an owned node, the expulsion step is treated separately from voluntary lazy movement. In the simulations reported in this paper, a forced non-owner agent is sent uniformly at random to one of the neighboring nodes of the owned node. At an endpoint of a path, it moves to the unique neighboring node. If directional bias is present, the agent resumes the directionally biased lazy random walk after this forced move.

This territorial rule stabilizes partial progress toward dispersion. Once a node has been claimed, its owner remains fixed there, and later arrivals cannot permanently crowd that node.

\subsection{Markov chain formulation and expected absorption time}

For a fixed graph $Q$, movement parameters, and behavioral rule, the process defines a finite Markov chain. In the baseline and directionally biased non-territorial models, the state space consists of population vectors. In the territorial models, the state space consists of augmented population--ownership states. In both cases, the dispersed configurations form the target set.

Although agents would continue to move under the physical movement rule, we stop the process once dispersion is achieved. Therefore, dispersed configurations are treated as absorbing states. Let $\mathcal{S}$ denote the finite state space and let $\mathcal{A}\subseteq\mathcal{S}$ denote the set of absorbing dispersed states. The transient state space is
\[
\mathcal{S}_0=\mathcal{S}\setminus\mathcal{A}.
\]
Suppose the transient states are indexed by $1,\dots,N$. Let
\[
B=(b_{ij})
\]
be the $N\times N$ transition matrix among transient states, where $b_{ij}$ is the probability of moving from transient state $i$ to transient state $j$ in one time step. The row sums of $B$ are at most one; the remaining probability mass corresponds to transitions from transient states into the absorbing set $\mathcal{A}$.

For each transient state $i$, let $T_i$ denote the expected number of steps required to reach a dispersed configuration when the process starts from state $i$. Conditioning on the first step gives
\[
T_i=1+\sum_{j=1}^{N}b_{ij}T_j,\qquad i=1,\dots,N.
\]
Writing
\[
T=(T_1,\dots,T_N)^{\mathsf T},
\]
the system can be expressed as
\[
T=\mathbf{1}_N+BT,
\]
or equivalently,
\[
(I_N-B)T=\mathbf{1}_N,
\]
where $\mathbf{1}_N$ is the $N$-dimensional vector of ones. When the absorbing set is reached with probability one from every transient state, $I_N-B$ is invertible and
\[
T=(I_N-B)^{-1}\mathbf{1}_N.
\]
The matrix
\[
F=(I_N-B)^{-1}
\]
is the fundamental matrix of the absorbing Markov chain.

For degenerate parameter choices under which the dispersed set cannot be reached from some transient states, the corresponding expected absorption time is infinite. In the optimization and numerical comparisons below, such degenerate parameter choices are excluded or treated as having infinite expected time.

This representation allows expected dispersion times to be computed for fixed movement parameters by constructing the transient transition matrix $B$ and solving the linear system above. In the exact calculations below, symmetries of paths and cycles are used to group equivalent states and reduce the size of the transition matrix. In the territorial cases, the reduced states shown in the figures represent aggregate population--ownership configurations. The transition probabilities in Section~III are computed from the movement and ownership rules described above.

\section{Three-Agent Dispersion on Small Networks}
\label{sec:three_agent}

We now specialize to the smallest nontrivial dispersion setting,
\[
m=n=3,\qquad D=1.
\]
The objective is to reach a dispersed configuration in which each of the three nodes contains exactly one agent. This section uses exact Markov-chain calculations to compare three movement rules: the baseline lazy random walk, the directionally biased lazy random walk without territorial behavior, and the territorial lazy random walk. The purpose of these small examples is not only to compute absorption times, but also to identify the local mechanisms that later appear in the larger-network simulations.

Two graph families are considered. The cycle $C_3$ has no boundary nodes and represents the simplest symmetric network. The path $L_3$ has two endpoints and one interior node, so movement at the boundary must be handled separately. Because these networks are small, the transition matrices and systems of equations can be written explicitly. We retain these calculations in the main text to make the absorption-time computation reproducible and to show directly how directional bias and territorial ownership change the underlying Markov chain.

\subsection{Cycle network $C_3$}

On the cycle $C_3$, the three nodes form a ring. With three agents and $D=1$, the absorbing configurations are exactly those in which each node contains one agent. Without territorial behavior, and up to rotation symmetry, the non-dispersed configurations can be grouped into states in which all agents occupy one node or two agents occupy one node while the third occupies another node. Alpern and Zeng~\cite{AlpernZeng2022} showed that, for the unbiased lazy random walk on this network, the loop random walk with $p=\frac{1}{3}$ minimizes the expected time to dispersion, and that the minimum expected time is $4.5$ steps, independently of the initial non-dispersed state.

\begin{figure}[!t]
\centering
\includegraphics[width=\columnwidth]{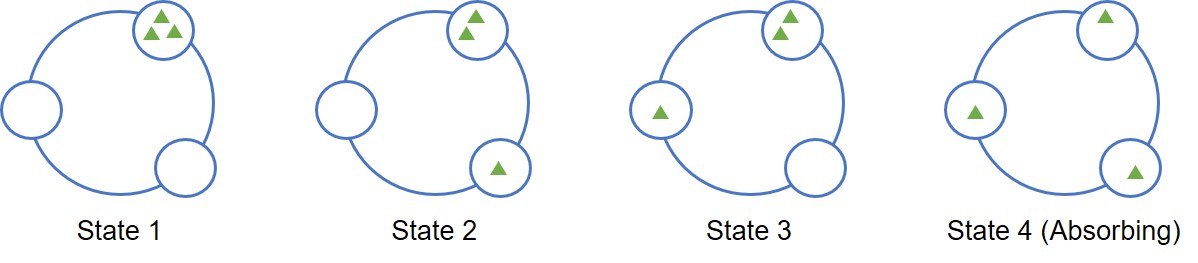}
\caption{Non-absorbing configurations for three-agent dispersion on $C_3$ under directionally biased lazy random walks, grouped up to rotation symmetry.}
\label{fig:C3_bias_states}
\end{figure}

We first examine whether directional bias improves this baseline result. Let $p$ denote the probability of staying at the current node, and let $r$ and $z$ denote the probabilities of moving clockwise and counterclockwise, respectively, with
\[
p+r+z=1.
\]
The non-absorbing configurations under directional bias are shown in Fig.~\ref{fig:C3_bias_states}. Using the ordering of states in the figure, the transient transition matrix is
\[
\resizebox{0.95\columnwidth}{!}{%
\ensuremath{\displaystyle
B=
\begin{pmatrix}
p^{3}+r^{3}+z^{3}        & 3p^{2}r+3r^{2}z+3pz^{2} & 3pr^{2}+3p^{2}z+3rz^{2} \\
pr^{2}+p^{2}z+rz^{2}    & p^{3}+r^{3}+6prz+z^{3}   & 3p^{2}r+3r^{2}z+3pz^{2} \\
p^{2}r+r^{2}z+pz^{2}    & 3pr^{2}+2p^{2}z+3rz^{2} & p^{3}+r^{3}+p^{2}z+6prz+z^{3}
\end{pmatrix}
}%
}
\]

For any valid choice of $(p,r,z)$, the expected absorption times are obtained from the fundamental matrix
\[
F=(I-B)^{-1}.
\]
The exact minimization shows that directional bias does not improve the optimal expected dispersion time on $C_3$. The minimum remains
$4.5$,
attained at the loop random walk
\[
p=r=z=\frac{1}{3}.
\]
Thus, on the smallest cycle, a shared direction of movement does not by itself create faster dispersion.

\begin{figure}[!t]
\centering
\includegraphics[width=0.75\columnwidth]{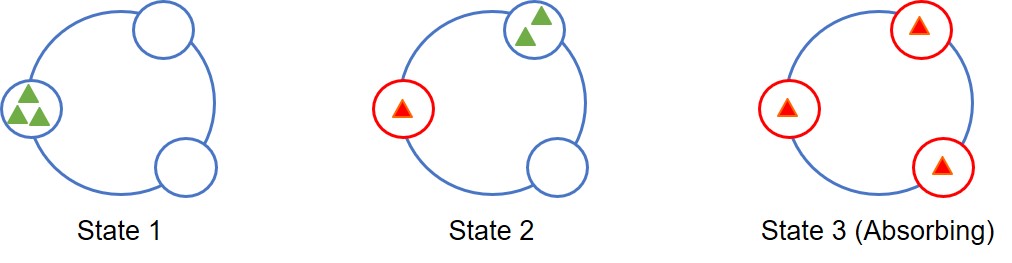}
\caption{Reduced non-absorbing configurations on $C_3$ with territorial behavior. Owned nodes and owners are indicated in red.}
\label{fig:C3terr1}
\end{figure}

We next introduce territorial behavior on $C_3$. Under territorial behavior, a node that contains exactly one agent after movement becomes owned by that agent, and later arrivals are forced to leave. The reduced non-absorbing states are shown in Fig.~\ref{fig:C3terr1}. State~1 represents the configuration in which all three agents occupy one unowned node. State~2 represents a configuration in which one node is owned and occupied by its owner, while the two remaining agents occupy another node.

\begin{figure}[!t]
\centering
\includegraphics[width=0.5\columnwidth]{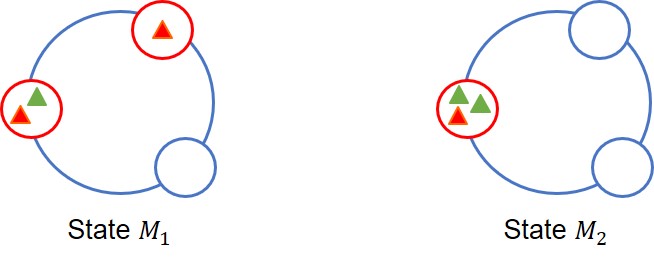}
\caption{Intermediate configurations reached from state~2 on $C_3$ with territorial behavior.}
\label{fig:C3terr2states1}
\end{figure}

Let $q=1-p$.
Starting from state~1, the process can remain in state~1, move to state~2, or reach dispersion in one step. Starting from state~2, the owner remains fixed and the two non-owner agents move according to the territorial rule. Two intermediate configurations, denoted by $M_1$ and $M_2$, are shown in Fig.~\ref{fig:C3terr2states1}.

Let $T_1$ and $T_2$ denote the expected dispersion times from states~1 and~2, respectively. Let $T_{M_1}$ and $T_{M_2}$ denote the corresponding expected times from the intermediate configurations. Direct enumeration of the territorial transitions gives
\[
\resizebox{0.85\columnwidth}{!}{%
\ensuremath{\displaystyle
\begin{aligned}
T_1
&=1+\left(p^{3}+\frac{q^{3}}{4}\right)T_1
+\left(3p^{2}q+\frac{3pq^{2}}{2}+\frac{3q^{3}}{4}\right)T_2,\\
T_2
&=1+\left(p^{2}+\frac{q^{2}}{4}\right)T_2
+\left(pq+\frac{q^{2}}{2}\right)T_{M_1}
+\frac{q^{2}}{4}T_{M_2},\\
T_{M_1}
&=1+\frac{1}{2}T_{M_1},\\
T_{M_2}
&=1+\frac{1}{2}T_2.
\end{aligned}
}%
}
\]

Solving this system gives
\[
\resizebox{0.75\columnwidth}{!}{%
\ensuremath{\displaystyle
\begin{aligned}
T_1(p)
&=
\frac{2\left(27p^{4}+18p^{3}-72p^{2}-16p-37\right)}
{3(p-1)(p+1)^{2}(11p+5)},\\
T_2(p)
&=
\frac{2\left(3p^{2}+2p-9\right)}
{(p-1)(11p+5)}.
\end{aligned}
}%
}
\]
The first expression is minimized at approximately
\[
p=0.37,\qquad T_1\approx 3.19,
\]
while the second is minimized at approximately
\[
p=0.42,\qquad T_2\approx 2.74.
\]
Compared with the non-territorial optimum $4.5$, territorial behavior reduces the expected dispersion time by about $29\%$ from state~1 and about $39\%$ from state~2.

This improvement has a simple interpretation. Once a node is claimed, the owner remains fixed and later arrivals cannot permanently occupy that node. Territorial behavior therefore prevents the process from repeatedly losing already achieved partial dispersion.

\begin{figure}[!t]
\centering
\includegraphics[width=0.9\columnwidth]{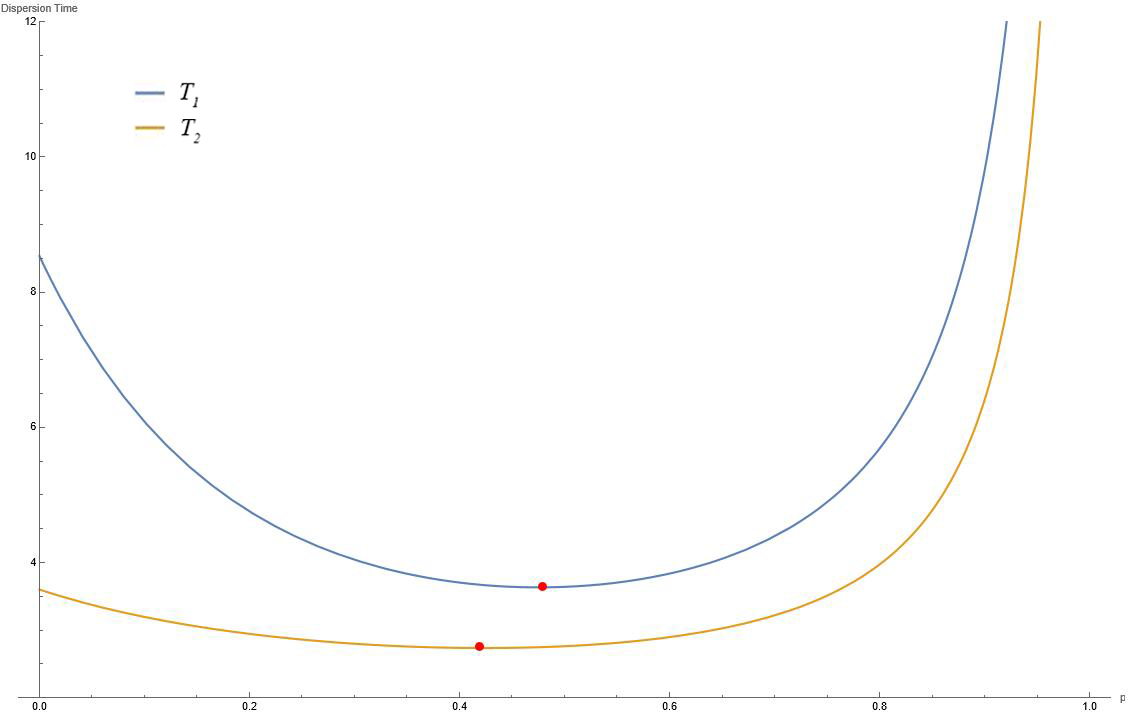}
\caption{Expected dispersion time on $C_3$ with territorial behavior as a function of the laziness parameter $p$.}
\label{fig:C3terr2time}
\end{figure}

\subsection{Path network $L_3$}

\begin{figure}[!t]
\centering
\includegraphics[width=0.9\columnwidth]{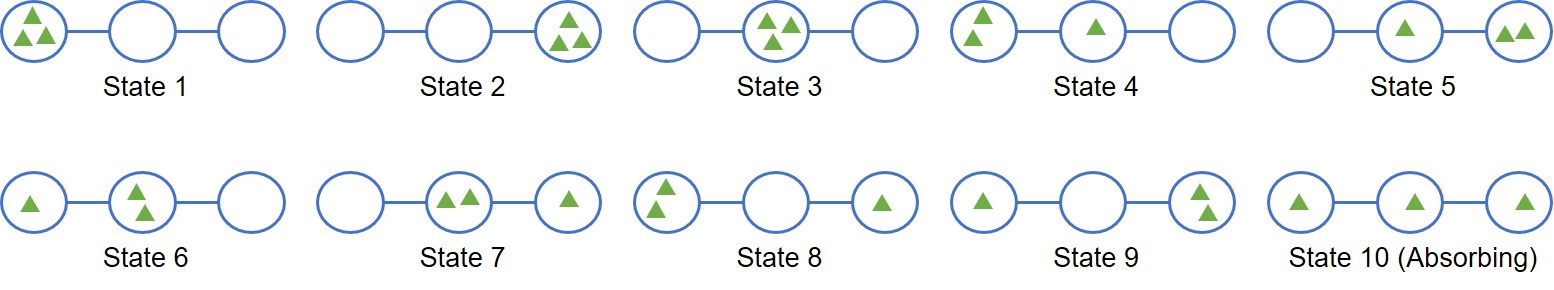}
\caption{Non-absorbing and absorbing configurations for three-agent dispersion on $L_3$.}
\label{fig:P3Biasterr1}
\end{figure}

The path $L_3$ has two endpoints and one interior node. This makes the transition structure more complex than on $C_3$, because movement at endpoints differs from movement at the interior node. With three agents and $D=1$, there are ten population configurations, one of which is dispersed. The nine non-absorbing configurations are shown in Fig.~\ref{fig:P3Biasterr1}.

We first consider the unbiased lazy random walk without territorial behavior. At an endpoint, an agent remains in place with probability $p$ and moves to the interior node with probability
\[
q=1-p.
\]
At the interior node, the agent remains in place with probability $p$ and moves to either endpoint with probability $q/2$. For completeness, we display the transient transition matrices used in the exact calculation. Using the state ordering in Fig.~\ref{fig:P3Biasterr1}, the transient transition matrix for the unbiased lazy random walk is
\[
\resizebox{0.98\columnwidth}{!}{%
\ensuremath{\displaystyle
B=
\begin{pmatrix}
p^{3}          & 0              & q^{3}         & 3p^{2}q     & 0              & 3pq^{2}     & 0              & 0              & 0 \\
0              & p^{3}          & q^{3}         & 0             & 3p^{2}q      & 0             & 3pq^{2}      & 0              & 0 \\
q^{3}/8      & q^{3}/8      & p^{3}         & 3pq^{2}/4 & 3pq^{2}/4  & 3p^{2}q/2 & 3p^{2}q/2  & 3q^{3}/8    & 3q^{3}/8 \\
p^{2}q/2    & 0              & pq^{2}       & p^{3}+pq^{2} & 0            & 2p^{2}q+q^{3}/2 & q^{3}/2  & p^{2}q/2    & 0 \\
0              & p^{2}q/2    & pq^{2}       & 0             & p^{3}+pq^{2} & q^{3}/2   & 2p^{2}q+q^{3}/2 & 0          & p^{2}q/2 \\
pq^{2}/4    & 0              & p^{2}q       & p^{2}q+q^{3}/4 & q^{3}/4 & p^{3}+pq^{2} & pq^{2}     & pq^{2}/2    & pq^{2}/4 \\
0              & pq^{2}/4    & p^{2}q       & q^{3}/4     & p^{2}q+q^{3}/4 & pq^{2}  & p^{3}+pq^{2} & pq^{2}/4 & pq^{2}/2 \\
0              & 0              & q^{3}         & p^{2}q       & 0              & 2pq^{2}     & pq^{2}        & p^{3}          & 0 \\
0              & 0              & q^{3}         & 0              & p^{2}q        & pq^{2}       & 2pq^{2}      & 0              & p^{3}
\end{pmatrix}
}%
}
\]

The directionally biased version changes the transition probabilities at the interior node. An agent at the interior node remains with probability $p$, moves right with probability $r$, and moves left with probability
\[
z=1-p-r.
\]
At endpoints, movement is unchanged: an agent stays with probability $p$ and moves to the unique neighbor with probability $q=1-p$. The corresponding transient transition matrix is
\[
\resizebox{0.98\columnwidth}{!}{%
\ensuremath{\displaystyle
B=
\begin{pmatrix}
p^{3}          & 0              & q^{3}         & 3p^{2}q          & 0                 & 3pq^{2}        & 0                 & 0              & 0 \\
0              & p^{3}          & q^{3}         & 0                  & 3p^{2}q         & 0                & 3pq^{2}         & 0              & 0 \\
z^{3}          & r^{3}          & p^{3}         & 3pz^{2}          & 3pr^{2}         & 3p^{2}z        & 3p^{2}r         & 3rz^{2}      & 3zr^{2} \\
p^{2}z        & 0              & pq^{2}       & p^{3}+2pzq    & 0                 & 2p^{2}q+zq^{2} & rq^{2}        & p^{2}r        & 0 \\
0              & p^{2}r        & pq^{2}       & 0                  & p^{3}+2prq   & zq^{2}          & 2p^{2}q+rq^{2} & 0            & p^{2}z \\
pz^{2}        & 0              & p^{2}q       & 2p^{2}z+qz^{2} & qr^{2}          & p^{3}+2pzq  & 2prq          & 2pzr       & pr^{2} \\
0              & r^{2}p        & p^{2}q       & z^{2}q            & 2p^{2}r+r^{2}q & 2pqz        & p^{3}+2pqr  & z^{2}p       & 2zrp \\
0              & 0              & q^{3}         & p^{2}q            & 0                 & 2pq^{2}        & q^{2}p          & p^{3}          & 0 \\
0              & 0              & q^{3}         & 0                  & p^{2}q           & q^{2}p          & 2pq^{2}        & 0              & p^{3}
\end{pmatrix}
}%
}
\]

For each initial state $i$, let $\hat{t}_i$ be the minimum expected dispersion time over the relevant parameter set. In the unbiased model, the optimization is over $p$. In the biased model, it is over $(p,r,z)$ subject to $p+r+z=1$. The exact results are shown in Table~\ref{tab:bias1}.

\begin{table}[!t]
\caption{Comparison of unbiased and directionally biased lazy random walks without territorial behavior on $L_3$.}
\label{tab:bias1}
\centering
\scriptsize
\resizebox{\columnwidth}{!}{%
\begin{tabular}{|c|cc||cccc|c|}
\hline
Initial & \multicolumn{2}{c||}{Unbiased} & \multicolumn{4}{c|}{Biased} & Decrease \\
state $i$ & $\hat{t}_i$ & $\hat{p}_i$ & $\hat{t}_i$ & $\hat{p}_i$ & $\hat{r}_i$ & $\hat{z}_i$ & in $\hat{t}_i$ \\
\hline
1 & 6.213 & 0.242 & 6.210 & 0.243 & 0.387 & 0.370 & $0.048\%$ \\
2 & 6.213 & 0.242 & 6.210 & 0.243 & 0.387 & 0.370 & $0.048\%$ \\
3 & 5.284 & 0.322 & 5.284 & 0.322 & 0.339 & 0.339 & $0\%$ \\
4 & 4.000 & 0     & 4.000 & 0     & 0.500 & 0.500 & $0\%$ \\
5 & 4.000 & 0     & 4.000 & 0     & 0.500 & 0.500 & $0\%$ \\
6 & 3.000 & 0     & 3.000 & 0     & 0.500 & 0.500 & $0\%$ \\
7 & 3.000 & 0     & 3.000 & 0     & 0.500 & 0.500 & $0\%$ \\
8 & 5.333 & 0.500 & 5.323 & 0.505 & 0.236 & 0.259 & $0.188\%$ \\
9 & 5.333 & 0.500 & 5.323 & 0.505 & 0.236 & 0.259 & $0.188\%$ \\
\hline
\end{tabular}%
}
\end{table}

Table~\ref{tab:bias1} shows that directional bias has negligible effect on $L_3$ when territorial behavior is absent. The largest reduction is only $0.188\%$, and several states show no improvement. Thus, even though directional bias changes the transition probabilities, it does not substantially change the optimal expected dispersion time on this smallest path network. Together with the calculation on $C_3$, this supports the conclusion that directional bias alone is not the main acceleration mechanism.

\begin{figure}[!t]
\centering
\includegraphics[width=\columnwidth]{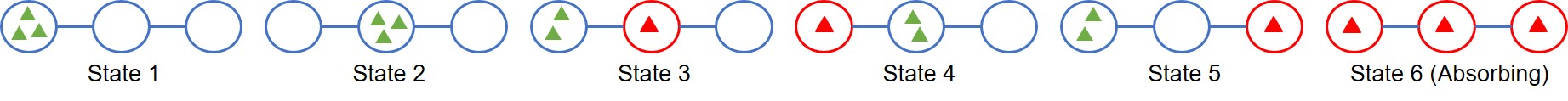}
\caption{Reduced non-absorbing configurations on $L_3$ with territorial behavior. Owned nodes and owners are indicated in red.}
\label{fig:P3terr1}
\end{figure}

We now add territorial behavior to the path $L_3$. Exploiting left-right symmetry, the non-absorbing configurations can be reduced to the five states shown in Fig.~\ref{fig:P3terr1}. Two reduced states contain no owned node, while the remaining reduced states contain one owned node and two non-owner agents distributed over the path.

\begin{figure}[!t]
\centering
\includegraphics[width=\columnwidth]{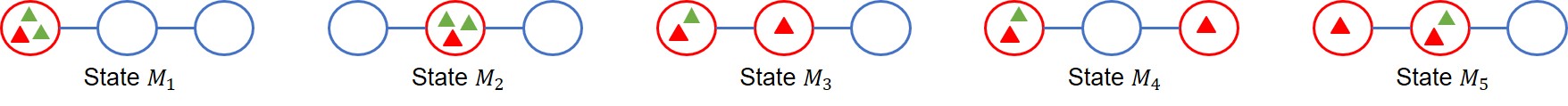}
\caption{Intermediate configurations on $L_3$ with territorial behavior.}
\label{fig:L3terr2}
\end{figure}

When non-owner agents enter an owned node, the territorial rule produces intermediate configurations before the process returns to one of the reduced states or reaches dispersion. These intermediate configurations are denoted by $M_1,\dots,M_5$ and are shown in Fig.~\ref{fig:L3terr2}.

Let $T_1,\dots,T_5$ denote the expected dispersion times from the five reduced states, and let $T_{M_1},\dots,T_{M_5}$ denote the expected times from the intermediate configurations. Direct enumeration of the territorial transitions gives
\[
\resizebox{0.98\columnwidth}{!}{%
\ensuremath{\displaystyle
\begin{aligned}
T_1
&=1+p^{3}T_1+q^{3}T_2+3p^{2}qT_3+3pq^{2}T_4,\\
T_2
&=1+2(q/2)^{3}T_1+p^{3}T_2+6p(q/2)^{2}T_3
+6p^{2}(q/2)T_4+6(q/2)^{3}T_5,\\
T_3
&=1+p^{2}T_3+q^{2}T_{M_2}+2pqT_{M_5},\\
T_4
&=1+p^{2}T_4+\frac{q^{2}}{4}T_5+\frac{q^{2}}{4}T_{M_1}
+pqT_{M_3}+\frac{q^{2}}{2}T_{M_4},\\
T_5
&=1+q^{2}T_4+p^{2}T_5,\\
T_{M_1}
&=1+T_4,\\
T_{M_2}
&=1+\frac{1}{2}T_3,\\
T_{M_3}
&=1+T_{M_5},\\
T_{M_4}
&=1,\\
T_{M_5}
&=1+\frac{1}{2}T_{M_3}.
\end{aligned}
}%
}
\]

Solving this system gives
\[
\resizebox{0.98\columnwidth}{!}{%
\ensuremath{\displaystyle
\begin{aligned}
T_1(p)
&=
-\frac{
46+258p+815p^2+1660p^3+785p^4-306p^5+93p^6-1308p^7-123p^8
}{
6(-1-11p-39p^2-43p^3+3p^4+3p^5+47p^6+35p^7+6p^8)
},\\
T_2(p)
&=
-\frac{
40+240p+605p^2+1006p^3+836p^4+108p^5+57p^6-858p^7-114p^8
}{
6(p-1)(1+8p+17p^2+6p^3)(1+4p+2p^2+4p^3+p^4)
},\\
T_3(p)
&=
\frac{2(5p^{2}-4p-2)}{3p^{2}-2p-1},\\
T_4(p)
&=
-\frac{8+16p-3p^{2}-13p^{3}}{2(p-1)(1+5p+2p^{2})},\\
T_5(p)
&=
-\frac{10+8p-23p^{2}+13p^{3}}{2(p-1)(1+5p+2p^{2})}.
\end{aligned}
}%
}
\]

When $p=0$,
the system simplifies to
\[
\footnotesize
T_1=6.75,\qquad
T_2=5.75,\qquad
T_3=4,\qquad
T_4=4,\qquad
T_5=5.
\]
For general $p\in(0,1)$, the minimum expected times are compared with the non-territorial case in Table~\ref{tab:terr2}.

\begin{table}[!t]
\caption{Unbiased lazy random walks with and without territorial behavior on $L_3$.}
\label{tab:terr2}
\centering
\begin{tabular}{|c|cc|c|}
\hline
Reduced & \multicolumn{2}{c|}{Minimum expected dispersion time} & Decrease \\
state   & Without territory & With territory & in time \\
\hline
1 & 6.21 & 4.62 & $25.6\%$ \\
2 & 5.28 & 3.74 & $29.2\%$ \\
3 & 4.00 & 4.00 & $0\%$ \\
4 & 3.00 & 3.25 & $-8.33\%$ \\
5 & 5.33 & 2.45 & $54.0\%$ \\
\hline
\end{tabular}
\end{table}

Table~\ref{tab:terr2} shows that territorial behavior usually reduces expected dispersion time on $L_3$, but not uniformly across all local configurations. The largest improvement occurs in reduced state~5, where the expected time decreases from $5.33$ to $2.45$, a reduction of $54.0\%$. Reduced state~4 is the only exception: the expected time increases from $3.00$ to $3.25$. This exception is useful because it shows that territorial behavior is not assumed to improve every local state. Its advantage comes from stabilizing partial progress, an effect that becomes clearer on larger networks.

\begin{figure}[!t]
\centering
\includegraphics[width=0.7\columnwidth]{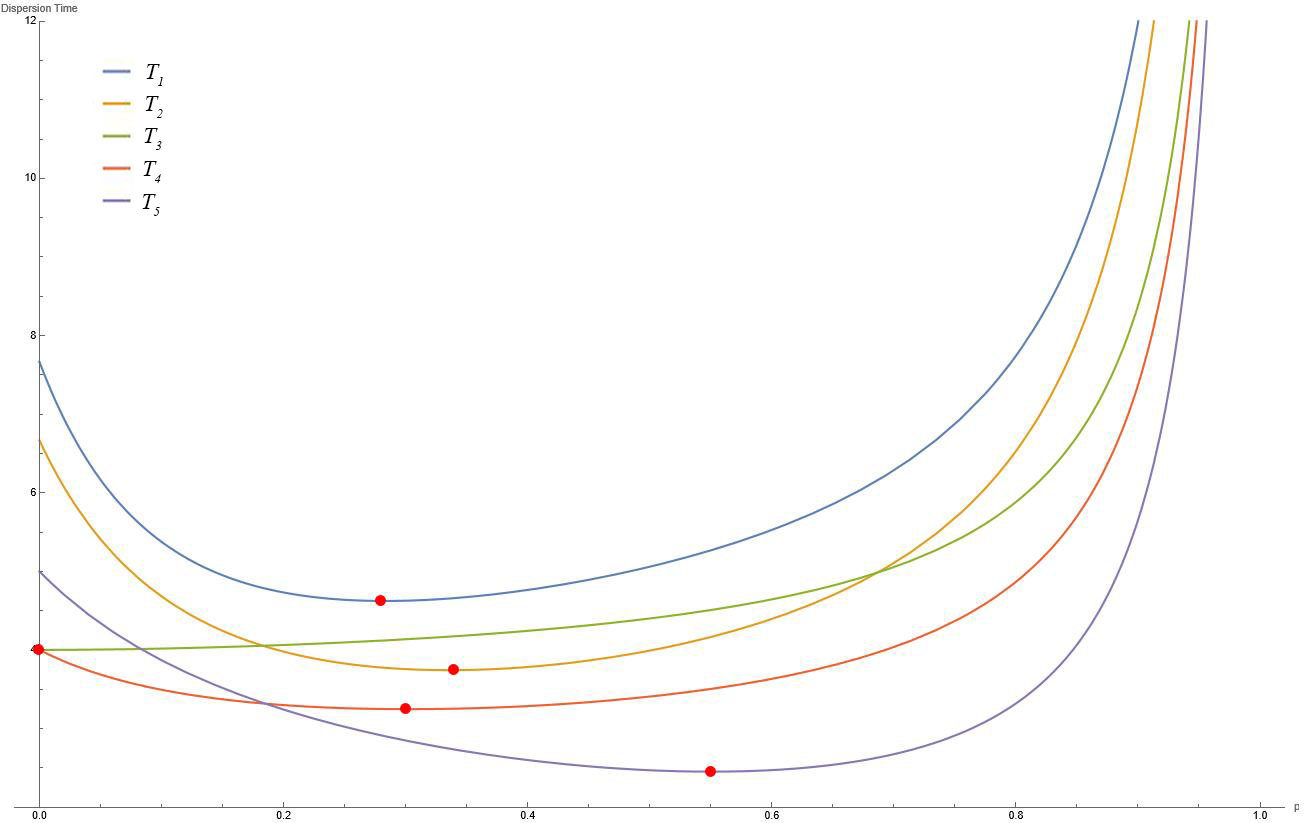}
\caption{Expected dispersion time on $L_3$ with territorial behavior for different reduced initial states as a function of $p$.}
\label{fig:L3terrtime}
\end{figure}

The exact calculations in this section lead to two conclusions. First, directional bias alone has little effect on the smallest path and cycle networks. On $C_3$, it does not improve the optimal expected dispersion time, and on $L_3$ the reductions are negligible. Second, territorial behavior can substantially reduce expected dispersion time because ownership stabilizes already occupied nodes. The effect is not uniformly positive for every local configuration, as shown by reduced state~4 on $L_3$, but the mechanism becomes increasingly important on larger networks. The next section therefore uses Monte Carlo simulation to examine how territorial behavior and directional bias scale on larger paths and cycles.

\section{Simulation Results on Larger Networks}
\label{sec:simulation}

The exact calculations in Section~III show how directional bias and territorial behavior affect dispersion on the smallest path and cycle networks. However, the number of population and ownership states grows rapidly with the number of nodes, making exact absorbing Markov-chain calculations impractical for larger instances. We therefore use Monte Carlo simulation to examine larger path and cycle networks.

The simulations have three objectives. First, we test whether the acceleration caused by territorial behavior persists as the network size increases. Second, we examine whether directional bias alone can substantially improve dispersion. Third, we study whether directional bias becomes more effective when combined with territorial behavior.

\subsection{Simulation setup and performance metric}
\label{subsec:simulation_setup}

All simulations in this section focus on the dispersion setting
\[
m=n,\qquad D=1.
\]
Thus, dispersion is achieved when every node contains exactly one agent. Each simulation run starts from a specified initial configuration. Time evolves in discrete steps, and all agents update their positions synchronously.

When territorial behavior is not active, each agent follows a lazy random walk. An agent at node $v$ remains at $v$ with probability $p$ and moves to a uniformly selected neighboring node with probability $1-p$. Hence each neighbor of $v$ is selected with probability
\[
\frac{1-p}{\deg(v)}.
\]

On path and cycle networks, we also consider directionally biased lazy random walks. At a node with two neighbors, the movement rule is parameterized by
$p+r+z=1$,
where the agent remains at its current node with probability $p$, moves to the right neighbor with probability $r$, and moves to the left neighbor with probability $z$. At an endpoint of a path, the agent remains at the endpoint with probability $p$ and moves to the unique neighboring node with probability $1-p$.

Territorial behavior is implemented using the ownership rule defined in Section~II. After each movement step, any unowned node containing exactly one agent becomes owned by that agent. Once an agent owns a node, it remains fixed at that node. If a non-owner agent is located at an owned node at the beginning of a movement step, it is forced to leave that node during that step. The forced move is treated as an expulsion step and is made uniformly at random among the neighboring nodes of the owned node. At an endpoint of a path, the forced non-owner moves to the unique neighboring node. If directional bias is present, the agent resumes the directionally biased lazy random walk after this forced displacement.

Let $S_t$ denote the state of the system after $t$ time steps, and let $d(S_t)$ be the minimum pairwise graph distance between agents. The stopping time is
\[
T=\inf\{t\geq 0: d(S_t)\geq D\}.
\]
Since this section considers only $D=1$, this is the first time at which all agents occupy distinct nodes.

For each graph, initial configuration, movement rule, and parameter setting, we perform $10000$
independent Monte Carlo trials. The expected dispersion time is estimated by the empirical mean of the stopping times. For each experiment, we numerically search over the admissible movement parameters and report the parameter values that give the smallest estimated mean dispersion time. In the unbiased case, the optimized parameter is denoted by $\hat{p}$. In the directionally biased case, the optimized parameters are denoted by $\hat{p}$, $\hat{r}$, and $\hat{z}$, and the corresponding optimized mean dispersion time is denoted by $\hat{t}$.

When comparing two movement rules, the percentage decrease in expected dispersion time is computed as
\[
\frac{\hat{t}_{\mathrm{baseline}}-\hat{t}_{\mathrm{modified}}}
{\hat{t}_{\mathrm{baseline}}}\times 100\%.
\]
A positive value means that the modified rule accelerates dispersion, while a negative value means that it is slower than the baseline rule for that initial configuration.

\subsection{Territorial behavior on cycles and paths}
\label{subsec:territorial_results}

We first isolate the effect of territorial behavior by comparing the baseline lazy random walk with the territorial lazy random walk. In this comparison, no directional bias is used. All agents are initially placed at the same node, and the objective is to disperse them across the entire network.

\begin{figure}[!t]
\centering
\subfloat[]{%
\includegraphics[width=\columnwidth]{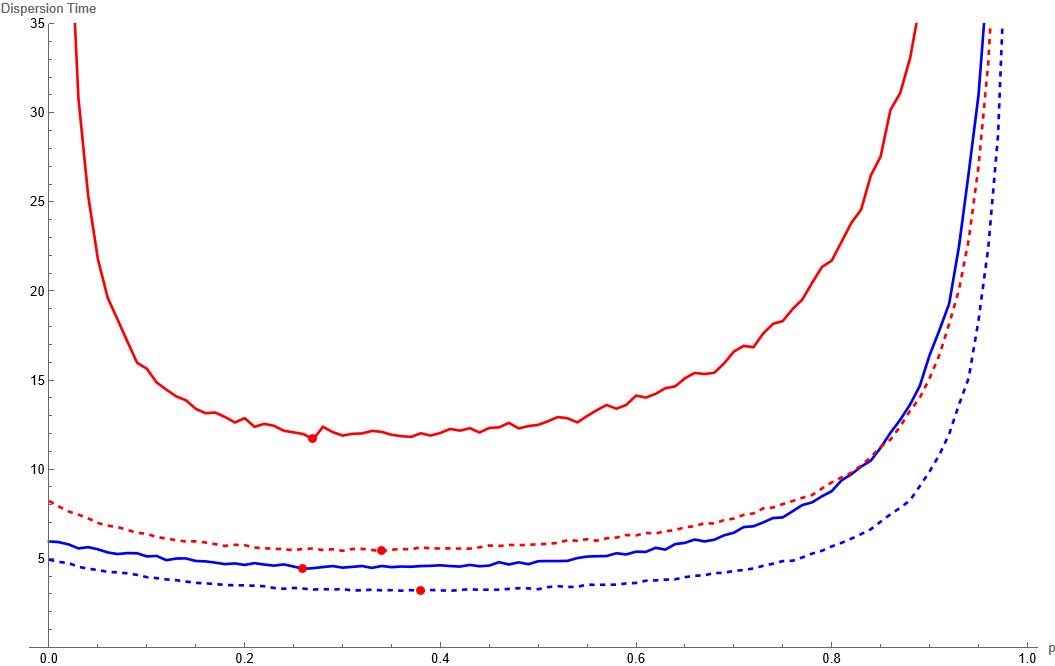}%
\label{fig:firstC}}
\\[-1mm]
\subfloat[]{%
\includegraphics[width=\columnwidth]{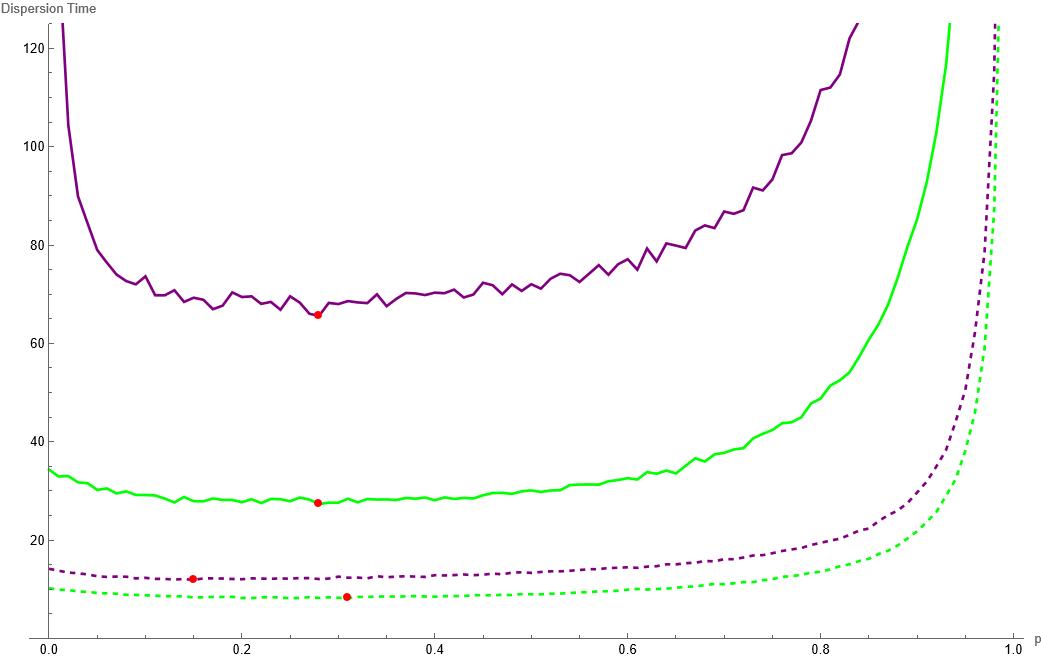}%
\label{fig:secondC}}
\caption{Expected dispersion time with and without territorial behavior on cycle networks. (a) $C_3$ and $C_4$. (b) $C_5$ and $C_6$.}
\label{fig:terrCircle}
\end{figure}

Fig.~\ref{fig:terrCircle} and Table~\ref{tab:terrCircle1} show the results for cycle networks. Territorial behavior substantially reduces the expected dispersion time, and the relative improvement increases with the size of the cycle. On $C_6$, the reduction is $81.8\%$. This supports the mechanism observed in Section~III: ownership stabilizes partial progress, because a claimed node is no longer repeatedly lost to crowding.

\begin{table}[!t]
\caption{Unbiased lazy random walks with and without territorial behavior on cycle networks.}
\label{tab:terrCircle1}
\centering
\begin{tabular}{|c|cc|cc|c|}
\hline
Cycle & \multicolumn{2}{c|}{Without territory} & \multicolumn{2}{c|}{With territory} & Decrease \\
      & Optimal $p$ & Time & Optimal $p$ & Time & in time \\
\hline
$C_3$ & 0.33 & 4.50 & 0.37 & 3.19 & $29.1\%$ \\
$C_4$ & 0.27 & 11.68 & 0.34 & 5.40  & $53.8\%$ \\
$C_5$ & 0.28 & 27.34 & 0.31 & 8.18  & $70.1\%$ \\
$C_6$ & 0.28 & 65.61 & 0.15 & 11.91 & $81.8\%$ \\
\hline
\end{tabular}
\end{table}

\begin{figure}[!t]
\centering
\subfloat[]{%
\includegraphics[width=\columnwidth]{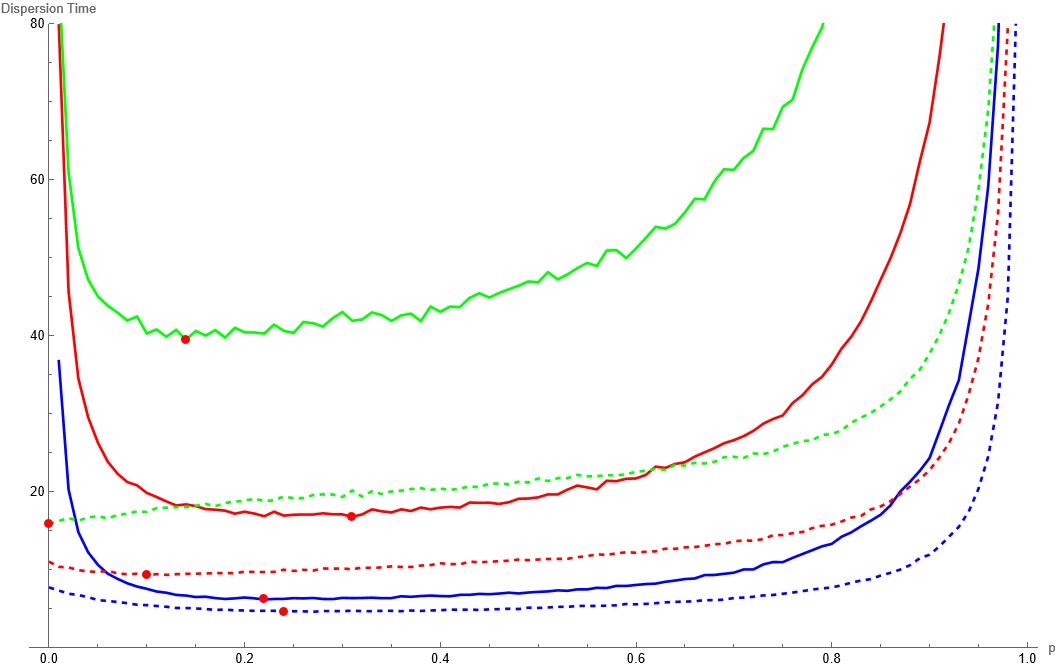}%
\label{fig:firstL}}
\\[-1mm]
\subfloat[]{%
\includegraphics[width=\columnwidth]{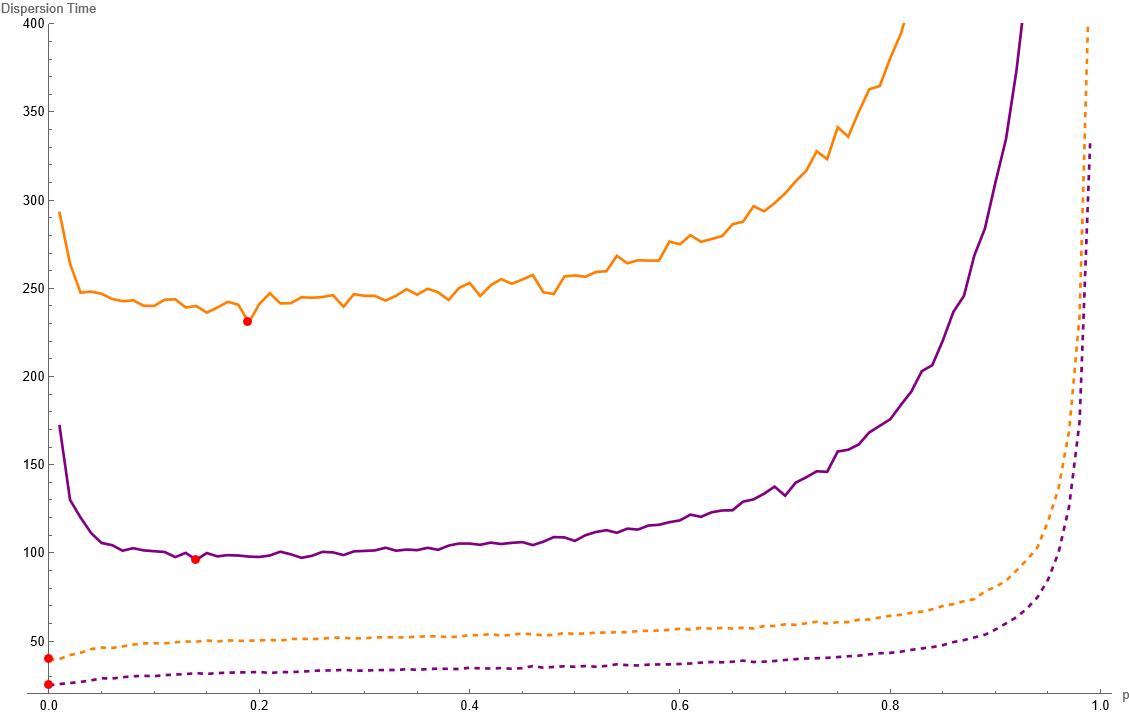}%
\label{fig:secondL}}
\caption{Expected dispersion time with and without territorial behavior on path networks. (a) $L_3$, $L_4$, and $L_5$. (b) $L_6$ and $L_7$.}
\label{fig:terrLine}
\end{figure}

The same pattern appears on path networks. Fig.~\ref{fig:terrLine} and Table~\ref{tab:terrLine1} show that territorial behavior also substantially reduces expected dispersion time on $L_n$. The reduction increases from $44.7\%$ on $L_4$ to $82.9\%$ on $L_7$. On paths, owned nodes act as anchors: once part of the path has been occupied by owners, the remaining non-owner agents are gradually pushed toward unclaimed locations.

\begin{table}[!t]
\caption{Unbiased lazy random walks with and without territorial behavior on path networks.}
\label{tab:terrLine1}
\centering
\begin{tabular}{|c|cc|cc|c|}
\hline
Path & \multicolumn{2}{c|}{Without territory} & \multicolumn{2}{c|}{With territory} & Decrease \\
     & Optimal $p$ & Time & Optimal $p$ & Time & in time \\
\hline
$L_4$ & 0.19 & 16.13  & 0.15 & 9.24  & $42.7\%$ \\
$L_5$ & 0.12 & 37.57  & 0.03 & 15.92 & $57.6\%$ \\
$L_6$ & 0.21 & 83.41  & 0    & 24.83 & $70.2\%$ \\
$L_7$ & 0.28 & 230.65 & 0    & 36.18 & $84.3\%$ \\
\hline
\end{tabular}
\end{table}

These simulations show that territorial behavior is the first major acceleration mechanism in the model. Its effect is visible in the exact calculations on $C_3$ and $L_3$, and it becomes stronger as the number of agents and nodes increases.

\subsection{Directional bias without territorial behavior}
\label{subsec:bias_without_territory}

We next examine directional bias in the absence of territorial behavior. The purpose is to determine whether a shared directional preference alone can accelerate dispersion. For cycle networks, the biased direction corresponds to clockwise or counterclockwise movement. For path networks, the bias applies at interior nodes, while endpoint movement is determined only by the probability of leaving the endpoint.

\subsubsection{Cycles $C_4$ and $C_5$}

\begin{figure}[!t]
\centering
\includegraphics[width=\columnwidth]{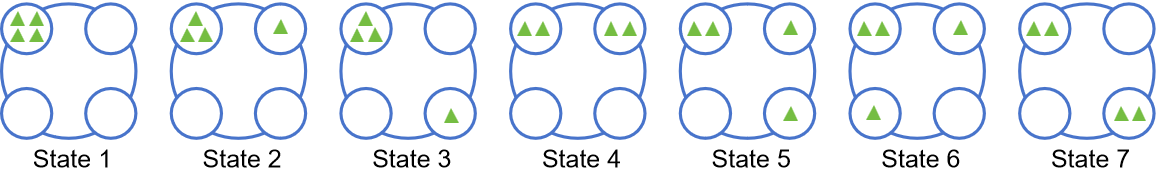}
\caption{Non-absorbing states in dispersion problems on $C_4$.}
\label{C4BiasState}
\end{figure}

\begin{figure}[!t]
\centering
\includegraphics[width=\columnwidth]{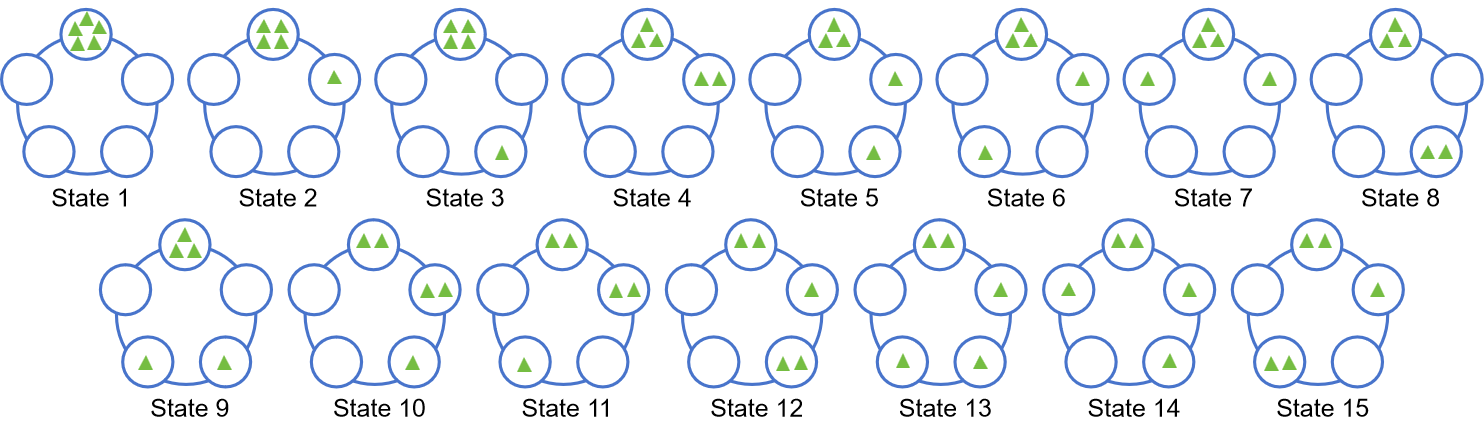}
\caption{Non-absorbing states in dispersion problems on $C_5$.}
\label{C5BiasState}
\end{figure}

The results for $C_4$ and $C_5$ are shown in Figs.~\ref{C4BiasState} and~\ref{C5BiasState} and Tables~\ref{tab:dispsimC4} and~\ref{tab:dispsimC5}. Overall, directional bias alone has only a limited effect on cycle networks. Most reductions are small, and some are negative, meaning that the optimized biased rule is slightly slower than the optimized unbiased rule for those initial states. A notable exception occurs on $C_4$, where one initial state has a reduction of $12.36\%$. This shows that directional bias can help in specific asymmetric configurations, but it does not provide a consistent acceleration mechanism by itself.

\begin{table}[!t]
\caption{Comparison of unbiased and directionally biased lazy random walks without territorial behavior on $C_4$.}
\label{tab:dispsimC4}
\centering
\scriptsize
\begin{tabular}{|c|cc||cccc|c|}
\hline
Initial & \multicolumn{2}{c||}{Unbiased} & \multicolumn{4}{c|}{Biased} & Decrease \\
state $i$ & $\hat{t}_i$ & $\hat{p}_i$ & $\hat{t}_i$ & $\hat{p}_i$ & $\hat{r}_i$ & $\hat{z}_i$ & in $\hat{t}_i$ \\
\hline
1 & 11.58 & 0.33 & 11.77 & 0.39 & 0.23 & 0.38 & $-1.64\%$ \\
2 & 10.97 & 0.29 & 10.58 & 0.35 & 0.36 & 0.29 & $3.56\%$ \\
3 & 10.83 & 0.48 & 10.71 & 0.43 & 0.31 & 0.26 & $1.11\%$ \\
4 & 3.97  & 0    & 3.97  & 0    & 0.50 & 0.50 & $0\%$ \\
5 & 10.58 & 0.39 & 10.25 & 0.48 & 0.17 & 0.35 & $3.12\%$ \\
6 & 4.08  & 0    & 3.90  & 0    & 0.48 & 0.52 & $4.41\%$ \\
7 & 10.52 & 0.40 & 9.22  & 0.53 & 0.46 & 0.01 & {\color[HTML]{FE0000}$12.36\%$} \\
\hline
\end{tabular}
\end{table}

\begin{table}[!t]
\caption{Comparison of unbiased and directionally biased lazy random walks without territorial behavior on $C_5$.}
\label{tab:dispsimC5}
\centering
\scriptsize
\begin{tabular}{|c|cc||cccc|c|}
\hline
Initial & \multicolumn{2}{c||}{Unbiased} & \multicolumn{4}{c|}{Biased} & Decrease \\
state $i$ & $\hat{t}_i$ & $\hat{p}_i$ & $\hat{t}_i$ & $\hat{p}_i$ & $\hat{r}_i$ & $\hat{z}_i$ & in $\hat{t}_i$ \\
\hline
1  & 27.47 & 0.26 & 26.85 & 0.25 & 0.43 & 0.32 & $2.26\%$ \\
2  & 27.20 & 0.20 & 26.95 & 0.29 & 0.33 & 0.38 & $0.92\%$ \\
3  & 27.08 & 0.20 & 26.63 & 0.20 & 0.38 & 0.42 & $1.66\%$ \\
4  & 27.40 & 0.20 & 26.83 & 0.33 & 0.33 & 0.34 & $2.08\%$ \\
5  & 25.95 & 0.22 & 26.11 & 0.30 & 0.44 & 0.26 & $-0.62\%$ \\
6  & 26.31 & 0.27 & 26.29 & 0.29 & 0.30 & 0.33 & $0.08\%$ \\
7  & 26.03 & 0.26 & 26.27 & 0.19 & 0.43 & 0.38 & $-0.92\%$ \\
8  & 25.61 & 0.30 & 24.91 & 0.40 & 0.33 & 0.27 & $2.73\%$ \\
9  & 25.65 & 0.39 & 25.53 & 0.34 & 0.31 & 0.35 & $0.47\%$ \\
10 & 24.44 & 0.01 & 24.65 & 0    & 0.54 & 0.46 & $-0.86\%$ \\
11 & 26.20 & 0.35 & 25.74 & 0.35 & 0.33 & 0.32 & $1.76\%$ \\
12 & 26.04 & 0.34 & 25.59 & 0.24 & 0.43 & 0.33 & $1.73\%$ \\
13 & 25.38 & 0.40 & 25.35 & 0.42 & 0.57 & 0.01 & $0.12\%$ \\
14 & 25.81 & 0.19 & 25.11 & 0.14 & 0.46 & 0.40 & $2.71\%$ \\
15 & 25.32 & 0.44 & 24.15 & 0.53 & 0.02 & 0.45 & $4.62\%$ \\
\hline
\end{tabular}
\end{table}

\subsubsection{Paths $L_4$ and $L_5$}

\begin{figure}[!t]
\centering
\includegraphics[width=\columnwidth]{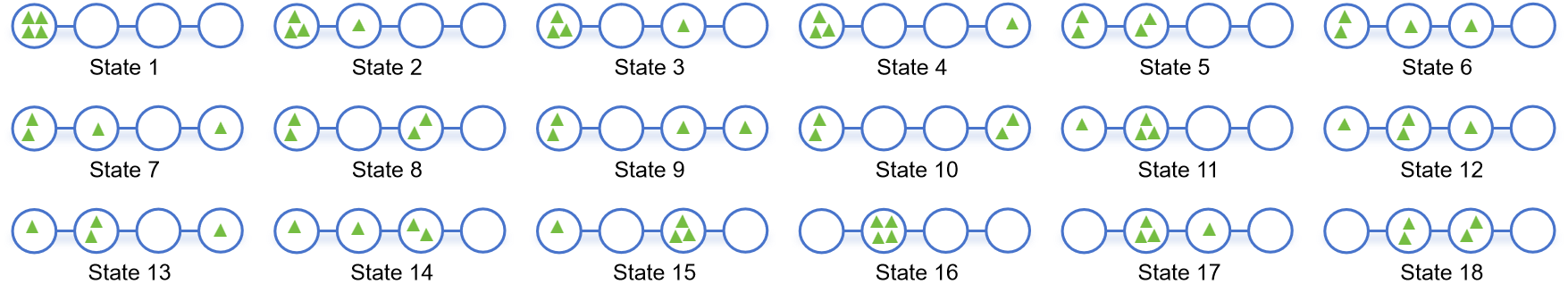}
\caption{Non-absorbing states in dispersion problems on $L_4$.}
\label{L4BiasState}
\end{figure}

\begin{figure}[!t]
\centering
\includegraphics[width=\columnwidth]{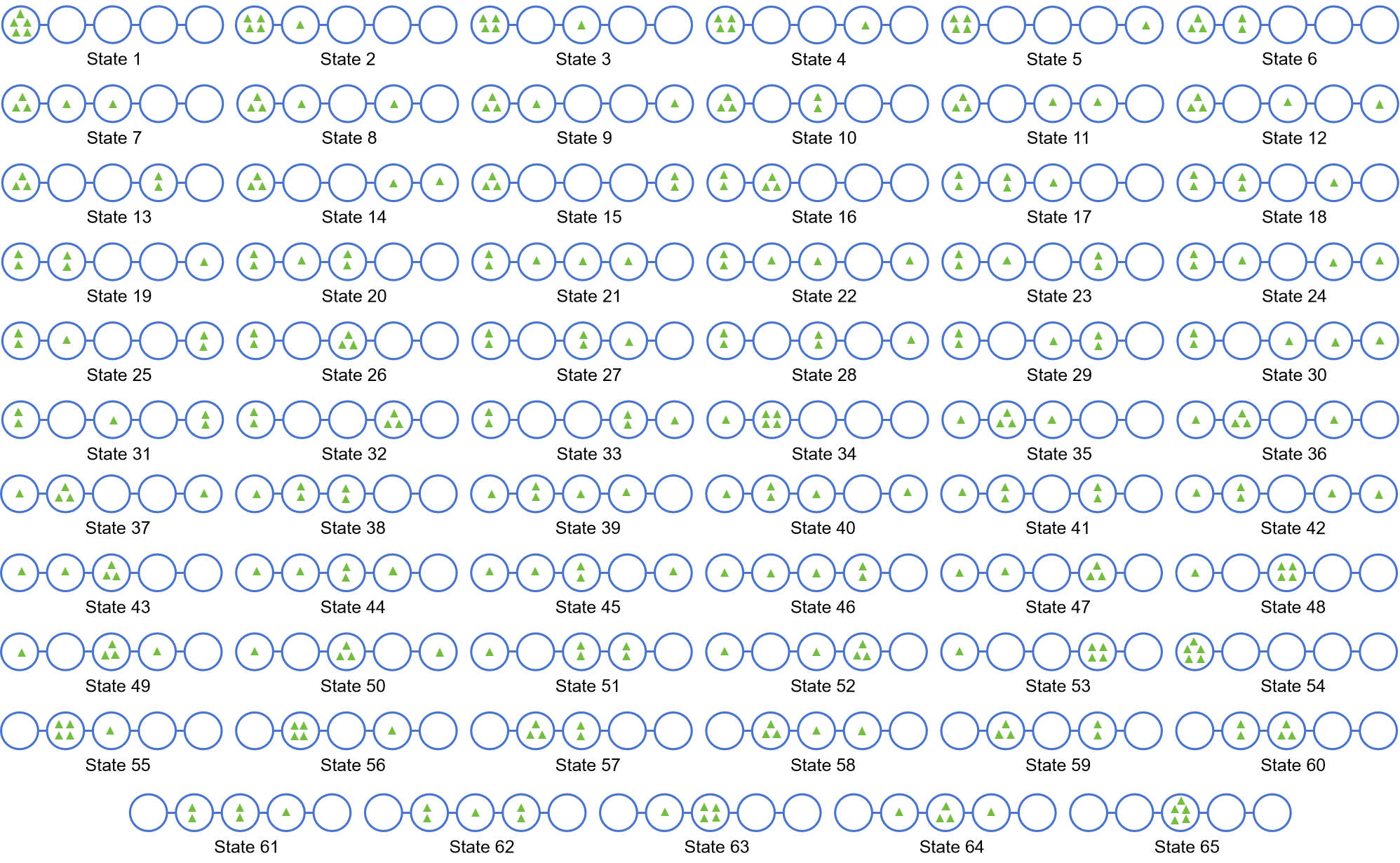}
\caption{Non-absorbing states in dispersion problems on $L_5$.}
\label{L5BiasState}
\end{figure}

The results for path networks $L_4$ and $L_5$ are shown in Figs.~\ref{L4BiasState} and~\ref{L5BiasState} and Tables~\ref{tab:dispsimL4} and~\ref{tab:dispsimL5}. One might expect directional bias to be more useful on paths because endpoints make the graph less symmetric. However, the results show that directional bias still has only a limited effect when territorial behavior is absent. Most reductions are close to zero, and some initial states become slightly slower under the biased rule.

\begin{table}[!t]
\caption{Comparison of unbiased and directionally biased lazy random walks without territorial behavior on $L_4$.}
\label{tab:dispsimL4}
\centering
\scriptsize
\begin{tabular}{|c|cc||cccc|c|}
\hline
Initial & \multicolumn{2}{c||}{Unbiased} & \multicolumn{4}{c|}{Biased} & Decrease \\
state $i$ & $\hat{t}_i$ & $\hat{p}_i$ & $\hat{t}_i$ & $\hat{p}_i$ & $\hat{r}_i$ & $\hat{z}_i$ & in $\hat{t}_i$ \\
\hline
1  & 16.89 & 0.21 & 16.52 & 0.26 & 0.39 & 0.35 & $2.19\%$ \\
2  & 16.04 & 0.21 & 16.12 & 0.23 & 0.39 & 0.38 & $-0.50\%$ \\
3  & 15.75 & 0.32 & 15.72 & 0.26 & 0.37 & 0.37 & $0.19\%$ \\
4  & 15.31 & 0.31 & 15.21 & 0.28 & 0.38 & 0.34 & $0.65\%$ \\
5  & 6.72  & 0    & 6.66  & 0    & 0.54 & 0.46 & $0.89\%$ \\
6  & 14.73 & 0.21 & 14.66 & 0.29 & 0.38 & 0.33 & $0.48\%$ \\
7  & 6.51  & 0    & 6.27  & 0    & 0.52 & 0.48 & $3.69\%$ \\
8  & 13.66 & 0.45 & 13.62 & 0.40 & 0.32 & 0.28 & $0.29\%$ \\
9  & 12.39 & 0.52 & 12.29 & 0.47 & 0.27 & 0.26 & $0.81\%$ \\
10 & 6.63  & 0    & 6.54  & 0    & 0.51 & 0.49 & $1.36\%$ \\
11 & 15.74 & 0.21 & 15.53 & 0.26 & 0.37 & 0.37 & $1.33\%$ \\
12 & 5.38  & 0    & 5.15  & 0    & 0.53 & 0.47 & $4.28\%$ \\
13 & 14.19 & 0.37 & 13.93 & 0.33 & 0.34 & 0.33 & $1.83\%$ \\
14 & 13.83 & 0.34 & 13.62 & 0.37 & 0.33 & 0.30 & $1.52\%$ \\
15 & 13.91 & 0.36 & 14.03 & 0.40 & 0.31 & 0.29 & $-0.86\%$ \\
16 & 15.66 & 0.28 & 15.45 & 0.26 & 0.38 & 0.36 & $1.34\%$ \\
17 & 14.11 & 0.26 & 14.16 & 0.26 & 0.39 & 0.35 & $-0.35\%$ \\
18 & 5.57  & 0    & 5.44  & 0    & 0.52 & 0.48 & $2.33\%$ \\
\hline
\end{tabular}
\end{table}

\begin{table}[!t]
\caption{Comparison of unbiased and directionally biased lazy random walks without territorial behavior on $L_5$.}
\label{tab:dispsimL5}
\centering
\scriptsize
\begin{tabular}{|c|cc||cccc|c|}
\hline
Initial & \multicolumn{2}{c||}{Unbiased} & \multicolumn{4}{c|}{Biased} & Decrease \\
state $i$ & $\hat{t}_i$ & $\hat{p}_i$ & $\hat{t}_i$ & $\hat{p}_i$ & $\hat{r}_i$ & $\hat{z}_i$ & in $\hat{t}_i$ \\
\hline
1  & 39.61 & 0.14 & 39.48 & 0.13 & 0.45 & 0.42 & $0.31\%$ \\
2  & 39.32 & 0.13 & 39.25 & 0.11 & 0.46 & 0.43 & $0.18\%$ \\
3  & 39.02 & 0.16 & 39.07 & 0.18 & 0.42 & 0.40 & $-0.14\%$ \\
4  & 38.58 & 0.11 & 38.46 & 0.22 & 0.39 & 0.39 & $0.31\%$ \\
5  & 37.97 & 0.21 & 38.07 & 0.17 & 0.43 & 0.40 & $-0.26\%$ \\
6  & 25.96 & 0    & 25.72 & 0    & 0.51 & 0.49 & $0.94\%$ \\
7  & 39.36 & 0.23 & 37.93 & 0.15 & 0.43 & 0.42 & $3.64\%$ \\
8  & 25.60 & 0    & 25.45 & 0    & 0.51 & 0.49 & $0.57\%$ \\
9  & 37.39 & 0.19 & 37.34 & 0.18 & 0.41 & 0.41 & $0.14\%$ \\
10 & 38.01 & 0.19 & 37.51 & 0.18 & 0.41 & 0.41 & $1.32\%$ \\
11 & 36.97 & 0.18 & 36.89 & 0.17 & 0.42 & 0.41 & $0.23\%$ \\
12 & 37.10 & 0.17 & 36.69 & 0.18 & 0.43 & 0.39 & $1.12\%$ \\
13 & 26.23 & 0    & 25.46 & 0    & 0.50 & 0.50 & $2.94\%$ \\
14 & 35.46 & 0.28 & 35.16 & 0.25 & 0.39 & 0.36 & $0.84\%$ \\
15 & 34.57 & 0.34 & 34.59 & 0.32 & 0.34 & 0.34 & $-0.05\%$ \\
16 & 25.02 & 0    & 25.03 & 0    & 0.49 & 0.51 & $-0.07\%$ \\
17 & 24.35 & 0    & 24.03 & 0    & 0.51 & 0.49 & $1.31\%$ \\
18 & 24.65 & 0    & 24.01 & 0    & 0.50 & 0.50 & $2.62\%$ \\
19 & 23.45 & 0    & 22.46 & 0    & 0.50 & 0.50 & $4.21\%$ \\
20 & 26.70 & 0.18 & 37.00 & 0.17 & 0.41 & 0.42 & {\color[HTML]{FE0000}$-38.58\%$} \\
21 & 23.49 & 0    & 23.45 & 0    & 0.52 & 0.48 & $0.18\%$ \\
22 & 36.43 & 0.27 & 35.79 & 0.26 & 0.38 & 0.36 & $1.76\%$ \\
23 & 23.99 & 0    & 24.40 & 0    & 0.50 & 0.50 & $-1.70\%$ \\
24 & 21.83 & 0    & 20.83 & 0    & 0.51 & 0.49 & $4.60\%$ \\
25 & 32.28 & 0.35 & 32.10 & 0.43 & 0.29 & 0.28 & $0.56\%$ \\
26 & 37.56 & 0.33 & 37.09 & 0.18 & 0.41 & 0.41 & $1.25\%$ \\
27 & 35.10 & 0.27 & 35.09 & 0.19 & 0.42 & 0.39 & $0.04\%$ \\
28 & 35.19 & 0.30 & 34.74 & 0.42 & 0.30 & 0.28 & $1.27\%$ \\
29 & 24.35 & 0    & 24.30 & 0    & 0.52 & 0.48 & $0.18\%$ \\
30 & 31.70 & 0.58 & 31.44 & 0.49 & 0.26 & 0.25 & $0.82\%$ \\
31 & 30.76 & 0.49 & 30.44 & 0.56 & 0.22 & 0.22 & $1.05\%$ \\
32 & 25.27 & 0    & 24.98 & 0    & 0.49 & 0.51 & $1.16\%$ \\
33 & 23.06 & 0    & 22.27 & 0    & 0.50 & 0.50 & $3.41\%$ \\
34 & 38.69 & 0.10 & 38.60 & 0.11 & 0.45 & 0.44 & $0.25\%$ \\
35 & 24.50 & 0    & 24.41 & 0    & 0.52 & 0.48 & $0.37\%$ \\
36 & 37.81 & 0.25 & 37.20 & 0.22 & 0.39 & 0.39 & $1.61\%$ \\
37 & 24.51 & 0    & 24.10 & 0    & 0.51 & 0.49 & $1.69\%$ \\
38 & 23.86 & 0    & 23.54 & 0    & 0.51 & 0.49 & $1.36\%$ \\
39 & 20.45 & 0    & 21.08 & 0    & 0.52 & 0.48 & $-3.09\%$ \\
40 & 23.37 & 0    & 22.99 & 0    & 0.51 & 0.49 & $1.63\%$ \\
41 & 34.69 & 0.24 & 33.83 & 0.24 & 0.39 & 0.37 & $2.46\%$ \\
42 & 18.00 & 0    & 17.48 & 0    & 0.52 & 0.48 & $2.92\%$ \\
43 & 37.01 & 0.30 & 36.43 & 0.22 & 0.39 & 0.39 & $1.56\%$ \\
44 & 22.94 & 0    & 22.25 & 0    & 0.49 & 0.51 & $3.02\%$ \\
45 & 34.70 & 0.17 & 34.89 & 0.21 & 0.40 & 0.39 & $-0.55\%$ \\
46 & 21.13 & 0    & 20.99 & 0    & 0.49 & 0.51 & $0.67\%$ \\
47 & 30.84 & 0.28 & 33.30 & 0.26 & 0.36 & 0.38 & {\color[HTML]{FE0000}$-7.98\%$} \\
48 & 36.08 & 0.28 & 36.50 & 0.22 & 0.39 & 0.39 & $-1.18\%$ \\
49 & 34.56 & 0.26 & 35.36 & 0.22 & 0.39 & 0.39 & $-2.33\%$ \\
50 & 35.57 & 0.38 & 34.59 & 0.36 & 0.32 & 0.32 & $2.77\%$ \\
51 & 24.04 & 0    & 23.51 & 0    & 0.49 & 0.51 & $2.21\%$ \\
52 & 24.04 & 0    & 24.64 & 0    & 0.49 & 0.51 & $-2.48\%$ \\
53 & 37.13 & 0.27 & 36.43 & 0.19 & 0.41 & 0.40 & $1.90\%$ \\
54 & 38.82 & 0.13 & 38.84 & 0.13 & 0.44 & 0.43 & $-0.07\%$ \\
55 & 37.87 & 0.17 & 37.44 & 0.18 & 0.41 & 0.41 & $1.14\%$ \\
56 & 37.69 & 0.31 & 37.02 & 0.25 & 0.38 & 0.37 & $1.78\%$ \\
57 & 25.01 & 0    & 24.52 & 0    & 0.50 & 0.50 & $1.96\%$ \\
58 & 35.40 & 0.25 & 35.29 & 0.25 & 0.38 & 0.37 & $0.31\%$ \\
59 & 34.20 & 0.35 & 34.71 & 0.35 & 0.32 & 0.33 & $-1.50\%$ \\
60 & 23.85 & 0    & 23.07 & 0    & 0.50 & 0.50 & $3.27\%$ \\
61 & 21.40 & 0    & 20.90 & 0    & 0.49 & 0.51 & $2.35\%$ \\
62 & 34.29 & 0.26 & 34.06 & 0.20 & 0.40 & 0.40 & $0.67\%$ \\
63 & 36.35 & 0.16 & 35.82 & 0.16 & 0.43 & 0.41 & $1.46\%$ \\
64 & 22.44 & 0    & 22.62 & 0    & 0.49 & 0.51 & $-0.78\%$ \\
65 & 36.55 & 0.28 & 36.62 & 0.24 & 0.38 & 0.38 & $-0.19\%$ \\
\hline
\end{tabular}
\end{table}

These results agree with the exact calculations in Section~III. Directional bias alone does not reliably stabilize partial progress toward dispersion. Even if agents tend to move in one direction, they can still revisit previously separated regions and create new collisions. Therefore, directional bias by itself is not the main source of acceleration.

\subsection{Directional bias with territorial behavior}
\label{subsec:bias_with_territory}

The role of directional bias changes when it is combined with territorial behavior. Territorial ownership creates stable occupied nodes, while directional bias can guide the remaining non-owner agents toward unclaimed parts of the network. We therefore examine whether these two mechanisms reinforce each other.

\subsubsection{Small paths and cycles with territorial behavior}

We first consider the same small path and cycle networks as above, but now with territorial behavior included. These experiments provide a bridge between the exact small-network calculations in Section~III and the larger owner-block simulations below. The purpose is to test whether directional bias becomes more useful once ownership is present.

Surprisingly, contrary to scenarios without territorial behaviour, introducing directional bias to a dispersion problem with territorial behaviour significantly accelerates the dispersion process in certain cases on path networks. With directional bias, there is a notable decrease on $L_4$, dispersion time decreases by $12\%$, $18\%$, $25\%$, $32\%$, $50\%$, $34\%$, $21\%$, and $61\%$ for initial states $1$, $2$, $8$, $9$, $10$, $11$, $12$, and $14$, respectively. Similarly, $38$ out of $65$, more than half of the initial states on $L_5$ experience significant decreases in the dispersion time, ranging from $10\%$ to $70\%$.

\begin{table}[!t]
\caption{Comparison of unbiased and directionally biased lazy random walks with territorial behavior on $L_4$.}
\label{tab:dispsimLTL4}
\centering
\scriptsize
\begin{tabular}{|c|cc||cccc|c|}
\hline
Initial & \multicolumn{2}{c||}{Unbiased} & \multicolumn{4}{c|}{Biased} & Decrease \\
state $i$ & $\hat{t}_i$ & $\hat{p}_i$ & $\hat{t}_i$ & $\hat{p}_i$ & $\hat{r}_i$ & $\hat{z}_i$ & in $\hat{t}_i$ \\
\hline
1  & 9.25 & 0.15 & 8.15 & 0.26 & 0.55 & 0.19 & {\color[HTML]{FE0000}$11.89\%$} \\
2  & 9.87 & 0.09 & 8.13 & 0.36 & 0.59 & 0.05 & {\color[HTML]{FE0000}$17.63\%$} \\
3  & 7.85 & 0.07 & 7.59 & 0.18 & 0.54 & 0.28 & $3.31\%$ \\
4  & 5.05 & 0.39 & 5.07 & 0.29 & 0.41 & 0.30 & $-0.40\%$ \\
5  & 6.38 & 0    & 6.00 & 0    & 0.59 & 0.41 & $5.96\%$ \\
6  & 8.37 & 0    & 7.60 & 0.40 & 0.59 & 0.01 & $9.20\%$ \\
7  & 4.45 & 0    & 4.16 & 0    & 0.39 & 0.61 & $6.52\%$ \\
8  & 6.39 & 0.30 & 4.77 & 0.71 & 0.28 & 0.01 & {\color[HTML]{FE0000}$25.35\%$} \\
9  & 2.91 & 0.67 & 1.98 & 0.50 & 0.01 & 0.49 & {\color[HTML]{FE0000}$31.96\%$} \\
10 & 4.36 & 0.62 & 2.20 & 0.55 & 0    & 0.45 & {\color[HTML]{FE0000}$49.54\%$} \\
11 & 7.81 & 0    & 5.13 & 0.42 & 0.57 & 0.01 & {\color[HTML]{FE0000}$34.31\%$} \\
12 & 5.22 & 0    & 4.10 & 0    & 0.66 & 0.34 & {\color[HTML]{FE0000}$21.46\%$} \\
13 & 3.28 & 0.29 & 3.05 & 0.70 & 0.29 & 0.01 & $7.01\%$ \\
14 & 5.42 & 0.41 & 2.13 & 0.50 & 0.49 & 0.01 & {\color[HTML]{FE0000}$60.70\%$} \\
15 & 4.79 & 0.30 & 4.46 & 0.30 & 0.49 & 0.21 & $6.89\%$ \\
16 & 8.19 & 0.13 & 7.59 & 0.20 & 0.52 & 0.28 & $7.33\%$ \\
17 & 7.09 & 0.10 & 6.99 & 0.08 & 0.48 & 0.44 & $1.41\%$ \\
18 & 4.86 & 0    & 4.86 & 0    & 0.52 & 0.48 & $0\%$ \\
\hline
\end{tabular}
\end{table}

On $L_5$, the effect of directional bias with territorial behavior is still configuration-dependent. Many states show only small improvements, while some states improve more clearly. This is consistent with the mechanism: directional bias helps only when the current ownership and population pattern creates a meaningful direction of progress.

\begin{table*}[!t]
\caption{Comparison of unbiased and directionally biased lazy random walks with territorial behavior on $L_5$.}
\label{tab:dispsimLTL5}
\centering
\scriptsize
\begin{tabular}{|c|cc||cccc|c|}
\hline
Initial & \multicolumn{2}{c||}{Unbiased} & \multicolumn{4}{c|}{Biased} & Decrease \\
state $i$ & $\hat{t}_i$ & $\hat{p}_i$ & $\hat{t}_i$ & $\hat{p}_i$ & $\hat{r}_i$ & $\hat{z}_i$ & in $\hat{t}_i$ \\
\hline
1  & 16.14 & 0.01 & 13.23 & 0.20 & 0.58 & 0.22 & {\color[HTML]{FE0000}$18.03\%$} \\
2  & 17.88 & 0    & 12.77 & 0.36 & 0.59 & 0.05 & {\color[HTML]{FE0000}$28.58\%$} \\
3  & 15.86 & 0.01 & 13.64 & 0.20 & 0.58 & 0.22 & {\color[HTML]{FE0000}$14.00\%$} \\
4  & 13.86 & 0.04 & 12.57 & 0.15 & 0.56 & 0.29 & $9.31\%$ \\
5  & 9.62  & 0.13 & 9.53  & 0.15 & 0.51 & 0.34 & $0.94\%$ \\
6  & 15.85 & 0.02 & 12.95 & 0.01 & 0.58 & 0.41 & {\color[HTML]{FE0000}$18.30\%$} \\
7  & 18.24 & 0.03 & 13.35 & 0.39 & 0.59 & 0.02 & {\color[HTML]{FE0000}$26.81\%$} \\
8  & 14.73 & 0.04 & 12.30 & 0.31 & 0.58 & 0.11 & {\color[HTML]{FE0000}$16.50\%$} \\
9  & 9.92  & 0.05 & 9.68  & 0.05 & 0.54 & 0.41 & $2.42\%$ \\
10 & 14.17 & 0.04 & 13.55 & 0.49 & 0.49 & 0.02 & $4.38\%$ \\
11 & 13.88 & 0.02 & 12.87 & 0.15 & 0.54 & 0.31 & $7.28\%$ \\
12 & 8.58  & 0.17 & 8.64  & 0.12 & 0.48 & 0.40 & $-0.70\%$ \\
13 & 10.44 & 0.27 & 7.46  & 0.60 & 0.39 & 0.01 & {\color[HTML]{FE0000}$28.54\%$} \\
14 & 5.61  & 0.40 & 5.44  & 0.26 & 0.36 & 0.38 & $3.03\%$ \\
15 & 7.44  & 0.58 & 6.98  & 0.60 & 0.39 & 0.01 & $6.18\%$ \\
16 & 16.43 & 0.02 & 12.72 & 0    & 0.59 & 0.41 & {\color[HTML]{FE0000}$22.58\%$} \\
17 & 16.04 & 0    & 13.12 & 0.01 & 0.57 & 0.42 & {\color[HTML]{FE0000}$18.20\%$} \\
18 & 12.98 & 0.02 & 11.64 & 0.02 & 0.56 & 0.42 & {\color[HTML]{FE0000}$10.32\%$} \\
19 & 7.07  & 0    & 7.23  & 0    & 0.51 & 0.49 & $-2.26\%$ \\
20 & 15.10 & 0.01 & 10.59 & 0.41 & 0.58 & 0.01 & {\color[HTML]{FE0000}$29.87\%$} \\
21 & 14.96 & 0.03 & 12.22 & 0.38 & 0.59 & 0.03 & {\color[HTML]{FE0000}$18.32\%$} \\
22 & 9.18  & 0.27 & 8.92  & 0.08 & 0.52 & 0.40 & $2.83\%$ \\
23 & 10.37 & 0.39 & 5.92  & 0.50 & 0.49 & 0.01 & {\color[HTML]{FE0000}$42.91\%$} \\
24 & 5.01  & 0.59 & 4.55  & 0    & 0.35 & 0.65 & $9.18\%$ \\
25 & 6.57  & 0.62 & 5.66  & 0.52 & 0.46 & 0.02 & {\color[HTML]{FE0000}$13.85\%$} \\
26 & 13.46 & 0.03 & 10.14 & 0.52 & 0.47 & 0.01 & {\color[HTML]{FE0000}$24.67\%$} \\
27 & 11.69 & 0.12 & 10.94 & 0.51 & 0.48 & 0.01 & $6.42\%$ \\
28 & 7.00  & 0.48 & 5.18  & 0.70 & 0.29 & 0.01 & {\color[HTML]{FE0000}$26.00\%$} \\
29 & 8.84  & 0.50 & 5.94  & 0.70 & 0.29 & 0.01 & {\color[HTML]{FE0000}$32.81\%$} \\
30 & 3.26  & 0.63 & 1.96  & 0.53 & 0    & 0.47 & {\color[HTML]{FE0000}$39.88\%$} \\
31 & 4.91  & 0.69 & 2.16  & 0.55 & 0    & 0.45 & {\color[HTML]{FE0000}$56.01\%$} \\
32 & 7.81  & 0.53 & 7.49  & 0.56 & 0.36 & 0.08 & $4.10\%$ \\
\hline
\end{tabular}
\hfill
\begin{tabular}{|c|cc||cccc|c|}
\hline
Initial & \multicolumn{2}{c||}{Unbiased} & \multicolumn{4}{c|}{Biased} & Decrease \\
state $i$ & $\hat{t}_i$ & $\hat{p}_i$ & $\hat{t}_i$ & $\hat{p}_i$ & $\hat{r}_i$ & $\hat{z}_i$ & in $\hat{t}_i$ \\
\hline
33 & 5.60  & 0.66 & 2.17  & 0.54 & 0    & 0.46 & {\color[HTML]{FE0000}$61.25\%$} \\
34 & 14.75 & 0    & 9.47  & 0.25 & 0.62 & 0.13 & {\color[HTML]{FE0000}$35.80\%$} \\
35 & 14.39 & 0    & 9.84  & 0.40 & 0.59 & 0.01 & {\color[HTML]{FE0000}$31.62\%$} \\
36 & 11.86 & 0    & 8.67  & 0.23 & 0.60 & 0.17 & {\color[HTML]{FE0000}$26.90\%$} \\
37 & 7.53  & 0    & 7.13  & 0.44 & 0.49 & 0.07 & $5.31\%$ \\
38 & 14.96 & 0.02 & 8.01  & 0    & 0.69 & 0.31 & {\color[HTML]{FE0000}$46.46\%$} \\
39 & 11.52 & 0    & 8.85  & 0.38 & 0.59 & 0.03 & {\color[HTML]{FE0000}$23.18\%$} \\
40 & 5.30  & 0    & 4.85  & 0    & 0.58 & 0.42 & $8.49\%$ \\
41 & 9.13  & 0.20 & 6.09  & 0.70 & 0.29 & 0.01 & {\color[HTML]{FE0000}$33.30\%$} \\
42 & 3.55  & 0.37 & 3.11  & 0.70 & 0.29 & 0.01 & {\color[HTML]{FE0000}$12.39\%$} \\
43 & 14.28 & 0.18 & 5.37  & 0.41 & 0.58 & 0.01 & {\color[HTML]{FE0000}$62.39\%$} \\
44 & 10.40 & 0    & 4.42  & 0.32 & 0.67 & 0.01 & {\color[HTML]{FE0000}$57.50\%$} \\
45 & 5.46  & 0.37 & 3.53  & 0.70 & 0.29 & 0.01 & {\color[HTML]{FE0000}$35.35\%$} \\
46 & 7.51  & 0.43 & 2.25  & 0.51 & 0.48 & 0.01 & {\color[HTML]{FE0000}$70.04\%$} \\
47 & 6.08  & 0.45 & 4.87  & 0.27 & 0.56 & 0.17 & {\color[HTML]{FE0000}$19.90\%$} \\
48 & 11.61 & 0.06 & 8.62  & 0.16 & 0.57 & 0.27 & {\color[HTML]{FE0000}$25.75\%$} \\
49 & 8.25  & 0    & 7.30  & 0    & 0.55 & 0.45 & {\color[HTML]{FE0000}$11.52\%$} \\
50 & 5.37  & 0.29 & 5.38  & 0.22 & 0.43 & 0.35 & $-0.19\%$ \\
51 & 7.02  & 0.04 & 5.85  & 0    & 0.59 & 0.41 & {\color[HTML]{FE0000}$16.67\%$} \\
52 & 8.16  & 0.24 & 8.09  & 0.39 & 0.43 & 0.18 & $0.86\%$ \\
53 & 8.87  & 0.20 & 9.02  & 0.15 & 0.47 & 0.38 & $-1.69\%$ \\
54 & 15.02 & 0.03 & 12.92 & 0.22 & 0.54 & 0.24 & {\color[HTML]{FE0000}$13.98\%$} \\
55 & 14.80 & 0.02 & 13.37 & 0.10 & 0.56 & 0.34 & $9.66\%$ \\
56 & 12.95 & 0.05 & 12.17 & 0.16 & 0.52 & 0.32 & $6.02\%$ \\
57 & 13.74 & 0.01 & 12.34 & 0.14 & 0.53 & 0.33 & {\color[HTML]{FE0000}$10.19\%$} \\
58 & 13.04 & 0.02 & 12.42 & 0.06 & 0.53 & 0.41 & $4.75\%$ \\
59 & 10.16 & 0.27 & 10.05 & 0.80 & 0.20 & 0    & $1.08\%$ \\
60 & 12.69 & 0.02 & 12.05 & 0.11 & 0.53 & 0.36 & $5.04\%$ \\
61 & 11.13 & 0    & 11.01 & 0.02 & 0.53 & 0.45 & $1.08\%$ \\
62 & 10.40 & 0.44 & 7.63  & 0.80 & 0.20 & 0    & {\color[HTML]{FE0000}$26.63\%$} \\
63 & 13.29 & 0.02 & 12.58 & 0.16 & 0.49 & 0.35 & $5.34\%$ \\
64 & 11.16 & 0    & 11.15 & 0    & 0.52 & 0.48 & $0.09\%$ \\
65 & 12.90 & 0.07 & 13.04 & 0.07 & 0.49 & 0.44 & $-1.09\%$ \\
\hline
\end{tabular}
\end{table*}

The cycle results show a similar pattern. On $C_4$ and $C_5$, directional bias with territorial behavior can substantially reduce dispersion time for some initial states, but the improvement is not uniform. This reflects the absence of endpoints on cycles: ownership can break symmetry locally, but the direction of progress depends on the current configuration.

\begin{table}[!t]
\caption{Comparison of unbiased and directionally biased lazy random walks with territorial behavior on $C_4$.}
\label{tab:dispsimCTC4}
\centering
\scriptsize
\begin{tabular}{|c|cc||cccc|c|}
\hline
Initial & \multicolumn{2}{c||}{Unbiased} & \multicolumn{4}{c|}{Biased} & Decrease \\
state $i$ & $\hat{t}_i$ & $\hat{p}_i$ & $\hat{t}_i$ & $\hat{p}_i$ & $\hat{r}_i$ & $\hat{z}_i$ & in $\hat{t}_i$ \\
\hline
1 & 5.40 & 0.37 & 5.09 & 0.43 & 0.17 & 0.40 & $5.74\%$ \\
2 & 4.83 & 0.32 & 4.38 & 0.48 & 0.04 & 0.48 & $9.32\%$ \\
3 & 4.13 & 0.37 & 3.88 & 0.39 & 0.26 & 0.35 & $6.05\%$ \\
4 & 3.93 & 0    & 3.99 & 0    & 0.45 & 0.55 & $-1.53\%$ \\
5 & 3.84 & 0.34 & 2.53 & 0.60 & 0.03 & 0.37 & {\color[HTML]{FE0000}$34.11\%$} \\
6 & 3.91 & 0.07 & 3.36 & 0.53 & 0.04 & 0.43 & {\color[HTML]{FE0000}$14.07\%$} \\
7 & 4.68 & 0.55 & 3.73 & 0.62 & 0    & 0.43 & {\color[HTML]{FE0000}$20.30\%$} \\
\hline
\end{tabular}
\end{table}

\begin{table}[!t]
\caption{Comparison of unbiased and directionally biased lazy random walks with territorial behavior on $C_5$.}
\label{tab:dispsimCTC5}
\centering
\scriptsize
\begin{tabular}{|c|cc||cccc|c|}
\hline
Initial & \multicolumn{2}{c||}{Unbiased} & \multicolumn{4}{c|}{Biased} & Decrease \\
state $i$ & $\hat{t}_i$ & $\hat{p}_i$ & $\hat{t}_i$ & $\hat{p}_i$ & $\hat{r}_i$ & $\hat{z}_i$ & in $\hat{t}_i$ \\
\hline
1  & 8.18 & 0.27 & 7.23 & 0.38 & 0.12 & 0.50 & {\color[HTML]{FE0000}$11.61\%$} \\
2  & 7.97 & 0.29 & 6.47 & 0.45 & 0.05 & 0.50 & {\color[HTML]{FE0000}$18.82\%$} \\
3  & 7.22 & 0.31 & 5.93 & 0.42 & 0.18 & 0.40 & {\color[HTML]{FE0000}$17.87\%$} \\
4  & 7.65 & 0.15 & 6.74 & 0.46 & 0.12 & 0.42 & {\color[HTML]{FE0000}$11.90\%$} \\
5  & 7.22 & 0.31 & 4.74 & 0.55 & 0.06 & 0.39 & {\color[HTML]{FE0000}$34.35\%$} \\
6  & 6.20 & 0.26 & 5.50 & 0.46 & 0.08 & 0.46 & {\color[HTML]{FE0000}$11.29\%$} \\
7  & 7.00 & 0.02 & 5.63 & 0.51 & 0.04 & 0.45 & {\color[HTML]{FE0000}$19.57\%$} \\
8  & 6.85 & 0.30 & 5.58 & 0.58 & 0.03 & 0.39 & {\color[HTML]{FE0000}$18.54\%$} \\
9  & 5.23 & 0.33 & 4.61 & 0.47 & 0.24 & 0.29 & {\color[HTML]{FE0000}$11.85\%$} \\
10 & 6.93 & 0.03 & 5.57 & 0.50 & 0.05 & 0.45 & {\color[HTML]{FE0000}$19.62\%$} \\
11 & 5.67 & 0    & 5.59 & 0.44 & 0.22 & 0.34 & $1.41\%$ \\
12 & 6.70 & 0.52 & 5.50 & 0.49 & 0.08 & 0.34 & {\color[HTML]{FE0000}$17.91\%$} \\
13 & 4.96 & 0.56 & 2.75 & 0.64 & 0.04 & 0.32 & {\color[HTML]{FE0000}$44.56\%$} \\
14 & 5.87 & 0.10 & 3.58 & 0.52 & 0.03 & 0.45 & {\color[HTML]{FE0000}$39.01\%$} \\
15 & 6.36 & 0.50 & 4.06 & 0.60 & 0.04 & 0.36 & {\color[HTML]{FE0000}$36.16\%$} \\
\hline
\end{tabular}
\end{table}

These small territorial simulations show why directional bias becomes more important in the larger experiments below. Directional bias alone does little because it does not preserve progress. Territorial behavior creates persistent owned nodes, and directional bias becomes useful once those owned nodes create an effective direction toward unclaimed locations.

\subsubsection{Path networks with consecutive territorial owners}

We next consider path networks in which the first $s$ nodes of $L_n$ are already owned by territorial agents and the remaining non-owner agents start from the next node. This creates a natural direction of progress: non-owner agents should move away from the owned block and toward the unclaimed part of the path. A rightward directional bias is therefore expected to help.

\begin{figure}[!t]
\centering
\includegraphics[width=\columnwidth]{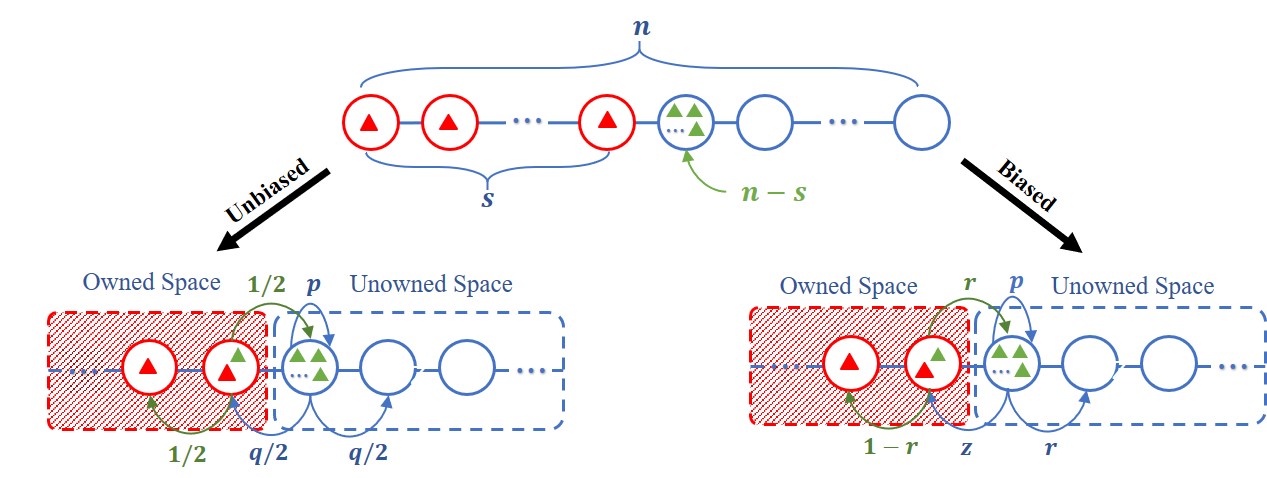}
\caption{Initial configurations with consecutive territorial owners on path networks.}
\label{Figure:OnwerBiased}
\end{figure}

Fig.~\ref{Figure:OnwerBiased} illustrates these initial configurations. Table~\ref{tab:OwnerLines} compares unbiased and directionally biased territorial lazy random walks on path networks with consecutive owners.

\begin{table*}[!t]
\caption{Comparison of unbiased and directionally biased lazy random walks with consecutive territorial owners on path networks. (a) $L_4$. (b) $L_5$. (c) $L_6$. (d) $L_7$. (e) $L_8$. (f) $L_9$. (g) $L_{10}$.}
\label{tab:OwnerLines}
\centering
\footnotesize
\renewcommand{\arraystretch}{0.95}

\subfloat[]{%
\begin{minipage}[t]{0.48\textwidth}
\centering
\begin{tabular*}{\linewidth}{@{\extracolsep{\fill}}|c|cc||cccc|c|}
\hline
$s$ & \multicolumn{2}{c||}{Unbiased} & \multicolumn{4}{c|}{Biased} & Decrease \\
    & $\hat{t}_s$ & $\hat{p}_s$ & $\hat{t}_s$ & $\hat{p}_s$ & $\hat{r}_s$ & $\hat{z}_s$ & in $\hat{t}_s$ \\
\hline
0 & 9.25 & 0.15 & 8.15 & 0.26 & 0.55 & 0.19 & {\color[HTML]{FE0000}$11.89\%$} \\
1 & 7.81 & 0    & 5.13 & 0.42 & 0.57 & 0.01 & {\color[HTML]{FE0000}$34.31\%$} \\
2 & 5.42 & 0.41 & 2.13 & 0.50 & 0.49 & 0.01 & {\color[HTML]{FE0000}$60.70\%$} \\
\hline
\end{tabular*}
\end{minipage}}
\hfil
\subfloat[]{%
\begin{minipage}[t]{0.48\textwidth}
\centering
\begin{tabular*}{\linewidth}{@{\extracolsep{\fill}}|c|cc||cccc|c|}
\hline
$s$ & \multicolumn{2}{c||}{Unbiased} & \multicolumn{4}{c|}{Biased} & Decrease \\
    & $\hat{t}_s$ & $\hat{p}_s$ & $\hat{t}_s$ & $\hat{p}_s$ & $\hat{r}_s$ & $\hat{z}_s$ & in $\hat{t}_s$ \\
\hline
0 & 16.14 & 0.01 & 13.23 & 0.20 & 0.58 & 0.22 & {\color[HTML]{FE0000}$18.03\%$} \\
1 & 14.75 & 0    & 9.47  & 0.25 & 0.62 & 0.13 & {\color[HTML]{FE0000}$35.80\%$} \\
2 & 14.28 & 0.18 & 5.37  & 0.41 & 0.58 & 0.01 & {\color[HTML]{FE0000}$62.39\%$} \\
3 & 7.51  & 0.43 & 2.25  & 0.51 & 0.48 & 0.01 & {\color[HTML]{FE0000}$70.04\%$} \\
\hline
\end{tabular*}
\end{minipage}}

\vspace{1mm}

\subfloat[]{%
\begin{minipage}[t]{0.48\textwidth}
\centering
\begin{tabular*}{\linewidth}{@{\extracolsep{\fill}}|c|cc||cccc|c|}
\hline
$s$ & \multicolumn{2}{c||}{Unbiased} & \multicolumn{4}{c|}{Biased} & Decrease \\
    & $\hat{t}_s$ & $\hat{p}_s$ & $\hat{t}_s$ & $\hat{p}_s$ & $\hat{r}_s$ & $\hat{z}_s$ & in $\hat{t}_s$ \\
\hline
0 & 24.94 & 0    & 19.59 & 0.20 & 0.60 & 0.20 & {\color[HTML]{FE0000}$21.45\%$} \\
1 & 24.70 & 0    & 14.64 & 0.23 & 0.62 & 0.15 & {\color[HTML]{FE0000}$40.73\%$} \\
2 & 25.73 & 0    & 10.30 & 0.39 & 0.59 & 0.02 & {\color[HTML]{FE0000}$59.97\%$} \\
3 & 20.16 & 0.24 & 5.63  & 0.40 & 0.59 & 0.01 & {\color[HTML]{FE0000}$72.07\%$} \\
4 & 9.71  & 0.50 & 2.38  & 0.50 & 0.49 & 0.01 & {\color[HTML]{FE0000}$75.49\%$} \\
\hline
\end{tabular*}
\end{minipage}}
\hfil
\subfloat[]{%
\begin{minipage}[t]{0.48\textwidth}
\centering
\begin{tabular*}{\linewidth}{@{\extracolsep{\fill}}|c|cc||cccc|c|}
\hline
$s$ & \multicolumn{2}{c||}{Unbiased} & \multicolumn{4}{c|}{Biased} & Decrease \\
    & $\hat{t}_s$ & $\hat{p}_s$ & $\hat{t}_s$ & $\hat{p}_s$ & $\hat{r}_s$ & $\hat{z}_s$ & in $\hat{t}_s$ \\
\hline
0 & 39.16 & 0    & 27.12 & 0.16 & 0.60 & 0.24 & {\color[HTML]{FE0000}$30.75\%$} \\
1 & 39.47 & 0    & 21.03 & 0.18 & 0.62 & 0.18 & {\color[HTML]{FE0000}$46.72\%$} \\
2 & 42.04 & 0    & 16.10 & 0.20 & 0.64 & 0.16 & {\color[HTML]{FE0000}$61.70\%$} \\
3 & 36.76 & 0.07 & 10.64 & 0.41 & 0.58 & 0.01 & {\color[HTML]{FE0000}$71.06\%$} \\
4 & 25.84 & 0.28 & 5.62  & 0.41 & 0.58 & 0.01 & {\color[HTML]{FE0000}$78.25\%$} \\
5 & 11.53 & 0.53 & 1.97  & 0.53 & 0.46 & 0.01 & {\color[HTML]{FE0000}$82.91\%$} \\
\hline
\end{tabular*}
\end{minipage}}

\vspace{1mm}

\subfloat[]{%
\begin{minipage}[t]{0.48\textwidth}
\centering
\begin{tabular*}{\linewidth}{@{\extracolsep{\fill}}|c|cc||cccc|c|}
\hline
$s$ & \multicolumn{2}{c||}{Unbiased} & \multicolumn{4}{c|}{Biased} & Decrease \\
    & $\hat{t}_s$ & $\hat{p}_s$ & $\hat{t}_s$ & $\hat{p}_s$ & $\hat{r}_s$ & $\hat{z}_s$ & in $\hat{t}_s$ \\
\hline
0 & 57.50 & 0    & 35.27 & 0.11 & 0.60 & 0.29 & {\color[HTML]{FE0000}$38.66\%$} \\
1 & 61.45 & 0    & 29.09 & 0.18 & 0.63 & 0.19 & {\color[HTML]{FE0000}$52.66\%$} \\
2 & 63.20 & 0    & 22.29 & 0.20 & 0.63 & 0.17 & {\color[HTML]{FE0000}$64.73\%$} \\
3 & 57.67 & 0.01 & 17.01 & 0.22 & 0.65 & 0.13 & {\color[HTML]{FE0000}$70.50\%$} \\
4 & 47.77 & 0.32 & 10.97 & 0.40 & 0.59 & 0.01 & {\color[HTML]{FE0000}$77.04\%$} \\
5 & 31.02 & 0.32 & 5.70  & 0.41 & 0.58 & 0.01 & {\color[HTML]{FE0000}$81.62\%$} \\
6 & 13.69 & 0.52 & 2.60  & 0.51 & 0.48 & 0.01 & {\color[HTML]{FE0000}$81.01\%$} \\
\hline
\end{tabular*}
\end{minipage}}
\hfil
\subfloat[]{%
\begin{minipage}[t]{0.48\textwidth}
\centering
\begin{tabular*}{\linewidth}{@{\extracolsep{\fill}}|c|cc||cccc|c|}
\hline
$s$ & \multicolumn{2}{c||}{Unbiased} & \multicolumn{4}{c|}{Biased} & Decrease \\
    & $\hat{t}_s$ & $\hat{p}_s$ & $\hat{t}_s$ & $\hat{p}_s$ & $\hat{r}_s$ & $\hat{z}_s$ & in $\hat{t}_s$ \\
\hline
0 & 80.18 & 0    & 45.22 & 0.12 & 0.61 & 0.27 & {\color[HTML]{FE0000}$43.60\%$} \\
1 & 86.73 & 0    & 37.22 & 0.14 & 0.61 & 0.25 & {\color[HTML]{FE0000}$57.09\%$} \\
2 & 89.43 & 0    & 30.52 & 0.20 & 0.63 & 0.17 & {\color[HTML]{FE0000}$65.87\%$} \\
3 & 84.91 & 0.01 & 24.34 & 0.21 & 0.64 & 0.15 & {\color[HTML]{FE0000}$71.33\%$} \\
4 & 73.39 & 0.02 & 17.64 & 0.22 & 0.66 & 0.12 & {\color[HTML]{FE0000}$75.96\%$} \\
5 & 57.87 & 0.01 & 11.06 & 0.41 & 0.58 & 0.01 & {\color[HTML]{FE0000}$80.89\%$} \\
6 & 37.42 & 0.38 & 5.86  & 0.40 & 0.59 & 0.01 & {\color[HTML]{FE0000}$84.34\%$} \\
7 & 15.48 & 0.66 & 2.59  & 0.51 & 0.48 & 0.01 & {\color[HTML]{FE0000}$83.27\%$} \\
\hline
\end{tabular*}
\end{minipage}}

\vspace{1mm}

\makebox[\textwidth][c]{%
\subfloat[]{%
\begin{minipage}[t]{0.60\textwidth}
\centering
\begin{tabular*}{\linewidth}{@{\extracolsep{\fill}}|c|cc||cccc|c|}
\hline
$s$ & \multicolumn{2}{c||}{Unbiased} & \multicolumn{4}{c|}{Biased} & Decrease \\
    & $\hat{t}_s$ & $\hat{p}_s$ & $\hat{t}_s$ & $\hat{p}_s$ & $\hat{r}_s$ & $\hat{z}_s$ & in $\hat{t}_s$ \\
\hline
0 & 113.60 & 0    & 54.98 & 0.13 & 0.59 & 0.28 & {\color[HTML]{FE0000}$51.60\%$} \\
1 & 117.28 & 0    & 45.98 & 0.14 & 0.62 & 0.24 & {\color[HTML]{FE0000}$60.79\%$} \\
2 & 122.41 & 0.01 & 38.04 & 0.14 & 0.63 & 0.23 & {\color[HTML]{FE0000}$68.92\%$} \\
3 & 117.75 & 0    & 32.37 & 0.20 & 0.63 & 0.17 & {\color[HTML]{FE0000}$72.51\%$} \\
4 & 105.47 & 0    & 25.05 & 0.21 & 0.64 & 0.15 & {\color[HTML]{FE0000}$76.25\%$} \\
5 & 87.41  & 0.02 & 18.68 & 0.39 & 0.59 & 0.02 & {\color[HTML]{FE0000}$78.63\%$} \\
6 & 68.65  & 0.11 & 11.29 & 0.41 & 0.58 & 0.01 & {\color[HTML]{FE0000}$83.55\%$} \\
7 & 42.78  & 0.65 & 5.77  & 0.40 & 0.59 & 0.01 & {\color[HTML]{FE0000}$86.51\%$} \\
8 & 17.39  & 0.60 & 2.78  & 0.51 & 0.48 & 0.01 & {\color[HTML]{FE0000}$84.01\%$} \\
\hline
\end{tabular*}
\end{minipage}}%
}
\end{table*}

The tables show that directional bias substantially reduces expected dispersion time when territorial owners already exist. The improvement generally becomes larger as the owned block grows. For small values of $s$, the best unbiased rule often has $\hat{p}=0$, meaning that non-owner agents should always move. As $s$ increases, the optimal stay probability often increases, indicating that it becomes useful for agents to remain at favorable positions after reaching them.

Under directional bias, the optimized rule usually assigns a high probability to rightward movement and a very small probability to leftward movement. Thus, on paths, territorial behavior and directional bias reinforce each other: territoriality stabilizes occupied nodes, and directional bias drives expansion into unoccupied territory.

\subsubsection{Path networks with all agents starting at node 1}

We now consider the harder case in which all agents start together at node~1. This corresponds to
$s=0$.
In this setting, no owned nodes are present initially. Territorial structure must first emerge through the movement process, after which directional bias may help extend the occupied region along the path.

\begin{table*}[!t]
\caption{Comparison of unbiased and directionally biased lazy random walks with territorial behavior on $L_n$ when $s=0$. (a) $n=15$--$40$. (b) $n=50$--$100$.}
\label{tab:dispsimLnN1}
\centering
\footnotesize
\renewcommand{\arraystretch}{0.95}

\subfloat[]{%
\begin{minipage}[t]{0.48\textwidth}
\centering
\begin{tabular*}{\linewidth}{@{\extracolsep{\fill}}|c|cc||cccc|c|}
\hline
$n$ & \multicolumn{2}{c||}{Unbiased} & \multicolumn{4}{c|}{Biased} & Decrease \\
    & $\hat{t}_i$ & $\hat{p}_i$ & $\hat{t}_i$ & $\hat{p}_i$ & $\hat{r}_i$ & $\hat{z}_i$ & in $\hat{t}_i$ \\
\hline
15 & 358.32  & 0    & 55.80  & 0.14 & 0.60 & 0.26 & {\color[HTML]{FE0000}$84.43\%$} \\
20 & 736.35  & 0    & 111.26 & 0.11 & 0.59 & 0.30 & {\color[HTML]{FE0000}$84.89\%$} \\
25 & 1265.94 & 0    & 160.44 & 0.08 & 0.60 & 0.32 & {\color[HTML]{FE0000}$87.33\%$} \\
30 & 1980.89 & 0    & 83.99  & 0.07 & 0.73 & 0.20 & {\color[HTML]{FE0000}$95.76\%$} \\
35 & 2840.48 & 0    & 94.33  & 0.08 & 0.74 & 0.19 & {\color[HTML]{FE0000}$96.68\%$} \\
40 & 3898.28 & 0.01 & 104.07 & 0.06 & 0.75 & 0.19 & {\color[HTML]{FE0000}$97.33\%$} \\
\hline
\end{tabular*}
\end{minipage}%
\label{tab:dispsimLnN1a}}
\hfil
\subfloat[]{%
\begin{minipage}[t]{0.48\textwidth}
\centering
\begin{tabular*}{\linewidth}{@{\extracolsep{\fill}}|c|cc||cccc|c|}
\hline
$n$ & \multicolumn{2}{c||}{Unbiased} & \multicolumn{4}{c|}{Biased} & Decrease \\
    & $\hat{t}_i$ & $\hat{p}_i$ & $\hat{t}_i$ & $\hat{p}_i$ & $\hat{r}_i$ & $\hat{z}_i$ & in $\hat{t}_i$ \\
\hline
50  & 6537.53  & 0.18 & 129.54 & 0.06 & 0.75 & 0.19 & {\color[HTML]{FE0000}$98.02\%$} \\
60  & 9921.33  & 0.22 & 154.46 & 0.04 & 0.76 & 0.20 & {\color[HTML]{FE0000}$98.44\%$} \\
70  & 14116.48 & 0.10 & 181.53 & 0.05 & 0.75 & 0.20 & {\color[HTML]{FE0000}$98.71\%$} \\
80  & 19045.42 & 0.07 & 202.16 & 0.03 & 0.76 & 0.21 & {\color[HTML]{FE0000}$98.94\%$} \\
90  & 24874.45 & 0.16 & 219.81 & 0.04 & 0.76 & 0.20 & {\color[HTML]{FE0000}$99.12\%$} \\
100 & 8093.64  & 0.07 & 203.68 & 0.20 & 0.15 & 0.65 & {\color[HTML]{FE0000}$97.48\%$} \\
\hline
\end{tabular*}
\end{minipage}%
\label{tab:dispsimLnN1b}}

\end{table*}

Table~\ref{tab:dispsimLnN1} shows that rightward directional bias greatly accelerates territorial dispersion on long paths. The reduction in mean dispersion time exceeds $80\%$ for $n\geq 15$,
and reaches $99.22\%$
on $L_{100}$. The optimized parameters show a clear pattern. In the unbiased territorial walk, the optimal value of $\hat{p}$ is often zero for smaller and medium path lengths, meaning that agents should keep moving. For larger paths, a small positive stay probability becomes useful. In contrast, under rightward bias, the optimal stay probability decreases as the path becomes longer, while the probability of moving right becomes dominant. This supports the interpretation that long-path dispersion is fastest when territorial ownership fixes progress and directional bias pushes the remaining agents toward unclaimed nodes.

\subsubsection{Cycle networks with consecutive territorial owners}

We now consider cycle networks. Cycles differ from paths because they have no endpoints. Therefore, a cycle does not have a permanent direction away from a boundary. However, once some consecutive nodes are owned, the ownership pattern creates an effective direction around the unclaimed part of the cycle.

\begin{figure}[!t]
\centering
\includegraphics[width=\columnwidth]{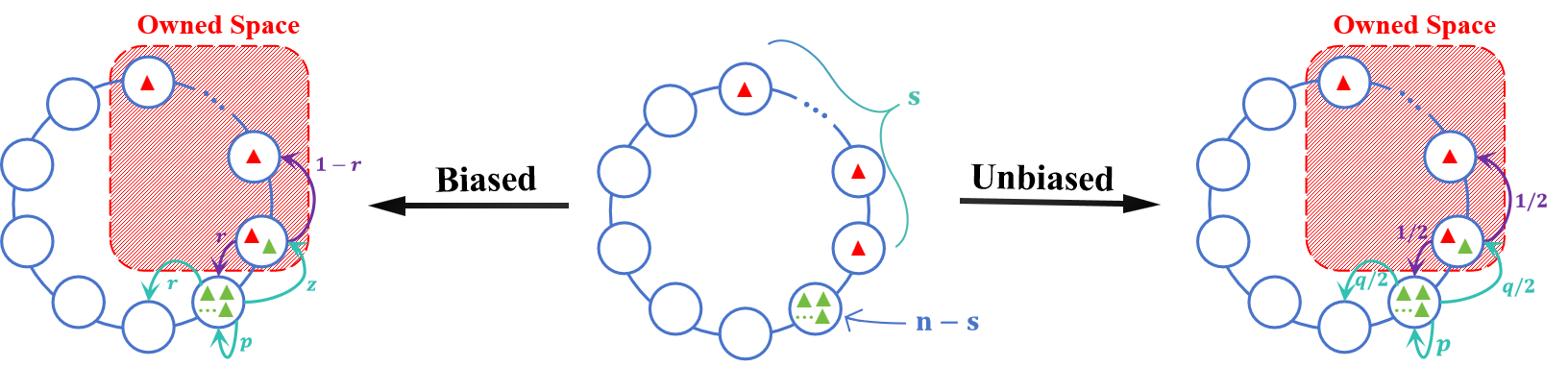}
\caption{Initial configurations with consecutive territorial owners on cycle networks.}
\label{Figure:OnwerCBiased}
\end{figure}

Fig.~\ref{Figure:OnwerCBiased} illustrates the initial configurations with consecutive territorial owners on cycle networks. Table~\ref{tab:OwnerCyclesCompare} compares unbiased and directionally biased territorial lazy random walks on $C_n$.

\begin{table*}[!t]
\caption{Comparison of unbiased and directionally biased lazy random walks with consecutive territorial owners on $C_n$, $n=4,\ldots,10$. (a) $n=4$. (b) $n=5$. (c) $n=6$. (d) $n=7$. (e) $n=8$. (f) $n=9$. (g) $n=10$.}
\label{tab:OwnerCyclesCompare}
\centering
\footnotesize
\renewcommand{\arraystretch}{0.95}

\subfloat[]{%
\begin{minipage}[t]{0.48\textwidth}
\centering
\begin{tabular*}{\linewidth}{@{\extracolsep{\fill}}|c|cc||cccc|c|}
\hline
$s$ & \multicolumn{2}{c||}{Unbiased} & \multicolumn{4}{c|}{Biased} & Decrease \\
    & $\hat{t}_s$ & $\hat{p}_s$ & $\hat{t}_s$ & $\hat{p}_s$ & $\hat{r}_s$ & $\hat{z}_s$ & in $\hat{t}_s$ \\
\hline
0 & 5.40 & 0.28 & 5.09 & 0.37 & 0.20 & 0.43 & $5.74\%$ \\
1 & 4.83 & 0.27 & 4.49 & 0.44 & 0.21 & 0.35 & $7.04\%$ \\
2 & 3.82 & 0.43 & 2.91 & 0.63 & 0.35 & 0.02 & {\color[HTML]{FE0000}$23.82\%$} \\
\hline
\end{tabular*}
\end{minipage}}
\hfil
\subfloat[]{%
\begin{minipage}[t]{0.48\textwidth}
\centering
\begin{tabular*}{\linewidth}{@{\extracolsep{\fill}}|c|cc||cccc|c|}
\hline
$s$ & \multicolumn{2}{c||}{Unbiased} & \multicolumn{4}{c|}{Biased} & Decrease \\
    & $\hat{t}_s$ & $\hat{p}_s$ & $\hat{t}_s$ & $\hat{p}_s$ & $\hat{r}_s$ & $\hat{z}_s$ & in $\hat{t}_s$ \\
\hline
0 & 8.19 & 0.24 & 7.20 & 0.37 & 0.16 & 0.47 & {\color[HTML]{FE0000}$12.09\%$} \\
1 & 7.94 & 0.27 & 6.70 & 0.44 & 0.16 & 0.40 & {\color[HTML]{FE0000}$15.62\%$} \\
2 & 7.34 & 0.38 & 5.94 & 0.51 & 0.15 & 0.34 & {\color[HTML]{FE0000}$19.07\%$} \\
3 & 4.90 & 0.54 & 3.16 & 0.74 & 0.25 & 0.01 & {\color[HTML]{FE0000}$35.51\%$} \\
\hline
\end{tabular*}
\end{minipage}}

\vspace{1mm}

\subfloat[]{%
\begin{minipage}[t]{0.48\textwidth}
\centering
\begin{tabular*}{\linewidth}{@{\extracolsep{\fill}}|c|cc||cccc|c|}
\hline
$s$ & \multicolumn{2}{c||}{Unbiased} & \multicolumn{4}{c|}{Biased} & Decrease \\
    & $\hat{t}_s$ & $\hat{p}_s$ & $\hat{t}_s$ & $\hat{p}_s$ & $\hat{r}_s$ & $\hat{z}_s$ & in $\hat{t}_s$ \\
\hline
0 & 11.93 & 0.15 & 9.36 & 0.38 & 0.11 & 0.51 & {\color[HTML]{FE0000}$21.54\%$} \\
1 & 11.91 & 0.20 & 8.82 & 0.46 & 0.09 & 0.45 & {\color[HTML]{FE0000}$25.94\%$} \\
2 & 11.39 & 0.25 & 8.13 & 0.52 & 0.08 & 0.40 & {\color[HTML]{FE0000}$28.62\%$} \\
3 & 9.89  & 0.29 & 7.26 & 0.59 & 0.10 & 0.31 & {\color[HTML]{FE0000}$26.59\%$} \\
4 & 6.08  & 0.50 & 3.30 & 0.72 & 0.27 & 0.01 & {\color[HTML]{FE0000}$45.72\%$} \\
\hline
\end{tabular*}
\end{minipage}}
\hfil
\subfloat[]{%
\begin{minipage}[t]{0.48\textwidth}
\centering
\begin{tabular*}{\linewidth}{@{\extracolsep{\fill}}|c|cc||cccc|c|}
\hline
$s$ & \multicolumn{2}{c||}{Unbiased} & \multicolumn{4}{c|}{Biased} & Decrease \\
    & $\hat{t}_s$ & $\hat{p}_s$ & $\hat{t}_s$ & $\hat{p}_s$ & $\hat{r}_s$ & $\hat{z}_s$ & in $\hat{t}_s$ \\
\hline
0 & 16.38 & 0.10 & 11.45 & 0.40 & 0.09 & 0.51 & {\color[HTML]{FE0000}$30.10\%$} \\
1 & 16.70 & 0.11 & 10.88 & 0.43 & 0.08 & 0.43 & {\color[HTML]{FE0000}$34.85\%$} \\
2 & 16.41 & 0.07 & 10.22 & 0.51 & 0.07 & 0.42 & {\color[HTML]{FE0000}$37.72\%$} \\
3 & 15.38 & 0.31 & 9.41  & 0.52 & 0.07 & 0.41 & {\color[HTML]{FE0000}$38.82\%$} \\
4 & 12.62 & 0.23 & 8.46  & 0.59 & 0.06 & 0.35 & {\color[HTML]{FE0000}$32.96\%$} \\
5 & 6.99  & 0.58 & 3.40  & 0.73 & 0.26 & 0.01 & {\color[HTML]{FE0000}$51.36\%$} \\
\hline
\end{tabular*}
\end{minipage}}

\vspace{1mm}

\subfloat[]{%
\begin{minipage}[t]{0.48\textwidth}
\centering
\begin{tabular*}{\linewidth}{@{\extracolsep{\fill}}|c|cc||cccc|c|}
\hline
$s$ & \multicolumn{2}{c||}{Unbiased} & \multicolumn{4}{c|}{Biased} & Decrease \\
    & $\hat{t}_s$ & $\hat{p}_s$ & $\hat{t}_s$ & $\hat{p}_s$ & $\hat{r}_s$ & $\hat{z}_s$ & in $\hat{t}_s$ \\
\hline
0 & 21.84 & 0.01 & 13.53 & 0.41 & 0.09 & 0.50 & {\color[HTML]{FE0000}$38.05\%$} \\
1 & 21.99 & 0.02 & 13.03 & 0.40 & 0.09 & 0.51 & {\color[HTML]{FE0000}$40.75\%$} \\
2 & 22.26 & 0    & 12.31 & 0.49 & 0.08 & 0.42 & {\color[HTML]{FE0000}$44.70\%$} \\
3 & 21.69 & 0.11 & 11.44 & 0.50 & 0.08 & 0.42 & {\color[HTML]{FE0000}$47.26\%$} \\
4 & 19.72 & 0.12 & 10.64 & 0.58 & 0.05 & 0.37 & {\color[HTML]{FE0000}$46.04\%$} \\
5 & 15.31 & 0.41 & 9.61  & 0.63 & 0.36 & 0.01 & {\color[HTML]{FE0000}$37.23\%$} \\
6 & 8.26  & 0.53 & 3.48  & 0.73 & 0.26 & 0.01 & {\color[HTML]{FE0000}$57.87\%$} \\
\hline
\end{tabular*}
\end{minipage}}
\hfil
\subfloat[]{%
\begin{minipage}[t]{0.48\textwidth}
\centering
\begin{tabular*}{\linewidth}{@{\extracolsep{\fill}}|c|cc||cccc|c|}
\hline
$s$ & \multicolumn{2}{c||}{Unbiased} & \multicolumn{4}{c|}{Biased} & Decrease \\
    & $\hat{t}_s$ & $\hat{p}_s$ & $\hat{t}_s$ & $\hat{p}_s$ & $\hat{r}_s$ & $\hat{z}_s$ & in $\hat{t}_s$ \\
\hline
0 & 27.34 & 0.01 & 15.63 & 0.41 & 0.08 & 0.51 & {\color[HTML]{FE0000}$42.83\%$} \\
1 & 28.20 & 0    & 15.12 & 0.41 & 0.08 & 0.51 & {\color[HTML]{FE0000}$46.38\%$} \\
2 & 28.60 & 0    & 14.43 & 0.45 & 0.09 & 0.46 & {\color[HTML]{FE0000}$49.55\%$} \\
3 & 28.71 & 0.01 & 13.63 & 0.48 & 0.08 & 0.44 & {\color[HTML]{FE0000}$52.53\%$} \\
4 & 27.30 & 0.04 & 12.77 & 0.57 & 0.05 & 0.38 & {\color[HTML]{FE0000}$53.22\%$} \\
5 & 24.29 & 0.25 & 11.90 & 0.57 & 0.06 & 0.37 & {\color[HTML]{FE0000}$51.01\%$} \\
6 & 17.86 & 0.31 & 10.32 & 0.60 & 0.39 & 0.01 & {\color[HTML]{FE0000}$42.22\%$} \\
7 & 9.32  & 0.65 & 3.56  & 0.70 & 0.29 & 0.01 & {\color[HTML]{FE0000}$61.80\%$} \\
\hline
\end{tabular*}
\end{minipage}}

\vspace{1mm}

\makebox[\textwidth][c]{%
\subfloat[]{%
\begin{minipage}[t]{0.60\textwidth}
\centering
\begin{tabular*}{\linewidth}{@{\extracolsep{\fill}}|c|cc||cccc|c|}
\hline
$s$ & \multicolumn{2}{c||}{Unbiased} & \multicolumn{4}{c|}{Biased} & Decrease \\
    & $\hat{t}_s$ & $\hat{p}_s$ & $\hat{t}_s$ & $\hat{p}_s$ & $\hat{r}_s$ & $\hat{z}_s$ & in $\hat{t}_s$ \\
\hline
0 & 34.19 & 0    & 17.72 & 0.44 & 0.08 & 0.48 & {\color[HTML]{FE0000}$48.17\%$} \\
1 & 35.11 & 0    & 17.24 & 0.44 & 0.09 & 0.47 & {\color[HTML]{FE0000}$50.90\%$} \\
2 & 36.89 & 0.01 & 16.52 & 0.47 & 0.08 & 0.45 & {\color[HTML]{FE0000}$55.22\%$} \\
3 & 36.60 & 0.01 & 15.74 & 0.45 & 0.08 & 0.47 & {\color[HTML]{FE0000}$56.99\%$} \\
4 & 35.56 & 0    & 14.97 & 0.48 & 0.07 & 0.45 & {\color[HTML]{FE0000}$57.90\%$} \\
5 & 33.44 & 0.08 & 14.06 & 0.56 & 0.05 & 0.39 & {\color[HTML]{FE0000}$57.95\%$} \\
6 & 28.86 & 0.13 & 13.10 & 0.61 & 0.06 & 0.33 & {\color[HTML]{FE0000}$54.61\%$} \\
7 & 20.95 & 0.39 & 11.35 & 0.60 & 0.39 & 0.01 & {\color[HTML]{FE0000}$45.82\%$} \\
8 & 10.38 & 0.60 & 3.60  & 0.74 & 0.25 & 0.01 & {\color[HTML]{FE0000}$65.32\%$} \\
\hline
\end{tabular*}
\end{minipage}}%
}

\end{table*}

Table~\ref{tab:OwnerCyclesCompare} shows that directional bias can reduce dispersion time on cycles with consecutive territorial owners, especially as the cycle size grows. The effect is more configuration-dependent than on paths. This is expected because a non-owner agent on a cycle can move around the owned region in either direction, whereas a path with an owned block has a clearer direction of progress.

Nevertheless, the simulations show that directional bias becomes useful on larger cycles. Once the cycle is large enough, the owned block breaks the symmetry of the network and creates a meaningful direction around the remaining unclaimed arc. Directional bias then helps non-owner agents avoid repeatedly returning to recently claimed nodes and instead move around the cycle toward unowned locations.

\subsubsection{Cycle networks with all agents starting at node 1}

Finally, we consider the case where all agents start together at node~1 on $C_n$. This again corresponds to
$s=0$.
Unlike the path case, there is no endpoint that defines a natural outward direction. Any benefit from directional bias must therefore come from the way ownership emerges and breaks the symmetry of the cycle.

\begin{table*}[!t]
\caption{Comparison of unbiased and directionally biased lazy random walks with territorial behavior on $C_n$ when $s=0$. (a) $n=10$--$35$. (b) $n=40$--$100$.}
\label{tab:dispsimCnN1}
\centering
\footnotesize
\renewcommand{\arraystretch}{0.95}

\subfloat[]{%
\begin{minipage}[t]{0.48\textwidth}
\centering
\begin{tabular*}{\linewidth}{@{\extracolsep{\fill}}|c|cc||cccc|c|}
\hline
$n$ & \multicolumn{2}{c||}{Unbiased} & \multicolumn{4}{c|}{Biased} & Decrease \\
    & $\hat{t}_i$ & $\hat{p}_i$ & $\hat{t}_i$ & $\hat{p}_i$ & $\hat{r}_i$ & $\hat{z}_i$ & in $\hat{t}_i$ \\
\hline
10 & 34.41  & 0 & 17.75 & 0.44 & 0.08 & 0.48 & {\color[HTML]{FE0000}$48.42\%$} \\
15 & 89.72  & 0 & 28.20 & 0.40 & 0.09 & 0.51 & {\color[HTML]{FE0000}$68.57\%$} \\
20 & 190.81 & 0 & 38.86 & 0.34 & 0.10 & 0.56 & {\color[HTML]{FE0000}$79.63\%$} \\
25 & 331.33 & 0 & 49.52 & 0.34 & 0.11 & 0.55 & {\color[HTML]{FE0000}$85.05\%$} \\
30 & 513.66 & 0 & 59.77 & 0.07 & 0.73 & 0.20 & {\color[HTML]{FE0000}$88.36\%$} \\
35 & 742.64 & 0 & 70.36 & 0.33 & 0.10 & 0.57 & {\color[HTML]{FE0000}$90.53\%$} \\
\hline
\end{tabular*}
\end{minipage}}
\hfil
\subfloat[]{%
\begin{minipage}[t]{0.48\textwidth}
\centering
\begin{tabular*}{\linewidth}{@{\extracolsep{\fill}}|c|cc||cccc|c|}
\hline
$n$ & \multicolumn{2}{c||}{Unbiased} & \multicolumn{4}{c|}{Biased} & Decrease \\
    & $\hat{t}_i$ & $\hat{p}_i$ & $\hat{t}_i$ & $\hat{p}_i$ & $\hat{r}_i$ & $\hat{z}_i$ & in $\hat{t}_i$ \\
\hline
40  & 1008.50 & 0    & 80.75  & 0.24 & 0.13 & 0.63 & {\color[HTML]{FE0000}$91.99\%$} \\
50  & 1695.46 & 0    & 101.61 & 0.24 & 0.14 & 0.62 & {\color[HTML]{FE0000}$94.01\%$} \\
60  & 2556.83 & 0    & 122.10 & 0.23 & 0.14 & 0.63 & {\color[HTML]{FE0000}$95.22\%$} \\
70  & 3634.48 & 0.09 & 142.84 & 0.20 & 0.14 & 0.66 & {\color[HTML]{FE0000}$96.07\%$} \\
80  & 4899.61 & 0    & 162.62 & 0.21 & 0.14 & 0.65 & {\color[HTML]{FE0000}$96.68\%$} \\
90  & 6370.56 & 0.05 & 183.12 & 0.21 & 0.14 & 0.65 & {\color[HTML]{FE0000}$97.13\%$} \\
100 & 8093.64 & 0.07 & 203.68 & 0.20 & 0.15 & 0.65 & {\color[HTML]{FE0000}$97.48\%$} \\
\hline
\end{tabular*}
\end{minipage}}

\end{table*}

Table~\ref{tab:dispsimCnN1} shows that directional bias can strongly accelerate territorial dispersion on large cycles even when all agents start together at one node. The reduction in mean dispersion time increases with network size and reaches
$97.48\%$
on $C_{100}$. In the unbiased territorial case, the optimal rule generally favors active movement, with small values of $\hat{p}$. Under directional bias, the optimized parameters consistently favor one direction around the cycle. In the node ordering used in the simulations, this appears as a leftward preference, with
$\hat{z}>\hat{r}$.
This differs from the path case, where rightward movement is preferred. The difference is not contradictory: on a path, rightward movement is defined relative to the endpoint and the owned block, whereas on a cycle the labels left and right depend on the chosen orientation of the ring.

\subsection{Summary of simulation findings}
\label{subsec:simulation_summary}

The simulations show three main patterns. First, territorial behavior alone substantially accelerates dispersion on both paths and cycles. The relative improvement increases with network size, indicating that local ownership becomes more valuable as more agents must separate.

Second, directional bias without territorial behavior has limited effect. In the small path and cycle instances tested, most improvements are small, and some initial states are slightly slower under the optimized biased walk. Thus, directional bias by itself does not reliably produce fast dispersion.

Third, directional bias becomes powerful when combined with territorial behavior. On paths, the combination creates a directed expansion process: owned nodes stabilize the occupied region, and directional bias pushes non-owner agents into unclaimed territory. On cycles, the effect is more configuration-dependent, but it can still be large when ownership creates an effective direction around the cycle.

Overall, the simulations support the main conclusion of the paper: simple local behavioral rules can greatly change the global time needed for decentralized agents to reach dispersed configurations. Territorial behavior is the primary acceleration mechanism, while directional bias is most effective when territorial ownership gives the preferred direction a clear operational meaning.

\section{Conclusion}
\label{sec:conclusion}

This paper studied decentralized network-agent dispersion under lazy random walks. We focused on the case $m=n$ and $D=1$, where the objective is to reach a configuration in which every node is occupied by exactly one agent. The baseline process was formulated as a finite absorbing Markov chain, and the expected absorption time was used to measure dispersion efficiency. We then compared this baseline with two local behavioral modifications: territorial behavior and directional bias.

The main conclusion is that territorial behavior is the strongest acceleration mechanism. By allowing an agent that is alone at a node to claim that node and repel later arrivals, the process preserves partial progress toward dispersion. This mechanism is already visible in the exact calculations on $C_3$ and $L_3$, and it becomes more pronounced in the Monte Carlo simulations on larger paths and cycles. Across these simulations, territorial behavior consistently reduces the expected dispersion time, with larger relative reductions as the network size increases.

Directional bias alone has a much weaker effect. In the exact calculations on $C_3$ and $L_3$, and in the small-network simulations without territorial behavior, allowing agents to share a preferred direction produces little or no improvement in most initial states. This shows that a directional preference by itself does not reliably prevent agents from revisiting already separated regions or creating new collisions.

The effect of directional bias changes when it is combined with territorial behavior. Once ownership creates a stable occupied region, directional bias can guide non-owner agents toward unclaimed nodes. This interaction is especially strong on paths, where an owned block creates a natural direction of progress. When all agents start at node~1, the reduction in expected dispersion time exceeds $80\%$ for $n\geq 15$ and reaches $99.22\%$ at $n=100$. On cycles, the effect is more dependent on the current ownership pattern because there is no boundary, but the combined rule still produces large reductions on larger networks, reaching $97.48\%$ at $n=100$ in the corresponding simulations.

These results show that simple local behavioral rules can substantially change the global time required for decentralized dispersion. In particular, improving dispersion does not only require increasing movement speed. A rule that stabilizes already occupied locations can be more important than movement rate alone, and directional bias is most useful when the current network configuration gives that direction a clear meaning.

Future work should extend the analysis beyond paths and cycles to graph classes with heterogeneous degree, bottlenecks, and community structure. Another direction is to study larger separation requirements $D>1$, where agents must avoid not only co-location but also nearby nodes. It would also be useful to consider agents with limited memory, partial observation, or limited communication, as well as strategic variants in which agents do not necessarily share the same objective.

\end{document}